\RequirePackage{fix-cm}

\documentclass[twocolumn,epjc3]{svjour3}  
\smartqed
\usepackage{amsmath}
\DeclareMathOperator\erfc{erfc}

\usepackage[titletoc,title]{appendix}
\usepackage{amssymb}
\usepackage{nccmath}
\usepackage{graphicx}
\usepackage{siunitx}
\DeclareSIUnit \photoelectron {PE}
\DeclareSIUnit \electron {e-}
\DeclareSIUnit \event {event}
\DeclareSIUnit \bar {bar}
\DeclareSIUnit \year {year}

\usepackage[normalem]{ulem}
\usepackage{ragged2e}
\usepackage{enumitem}
\usepackage{graphicx}
\usepackage{float}
\usepackage{hyperref}
\usepackage{multirow}
\usepackage[nolist,nohyperlinks,printonlyused]{acronym}
\usepackage{url}

\usepackage[switch]{lineno} 

\journalname{Eur. Phys. J. C}

\makeatletter
\newcommand*{\centerfloat}{%
  \parindent \z@
  \leftskip \z@ \@plus 1fil \@minus \textwidth
  \rightskip\leftskip
  \parfillskip \z@skip}
\makeatother

\begin{document}
\title{Measurement of the \acl{XA}'s Absolute \acl{PDE} for the \acl{DUNE}'s \acl{VD} \acl{FD}}

\author{
G.~Botogoske\thanksref{e2,addr3,addr4}
\and
F.~Bruni\thanksref{addr5}
\and
E.~Calvo\thanksref{addr1}
\and
R.~Calabrese\thanksref{addr2,addr3}
\and
N.~Canci\thanksref{addr3}
\and
A.~Canto\thanksref{addr1}
\and
C.M.~Cattadori\thanksref{addr5}
\and
A.~Cervera Villanueva\thanksref{addr8}
\and
S.~Coleman\thanksref{addr9}
\and
J.I.~Crespo-Anad\'{o}n\thanksref{addr1}
\and
C.~Cuesta\thanksref{addr1}
\and
F.~Di Capua\thanksref{addr2, addr3}
\and
N.~Durand\thanksref{addr9}
\and
G. Fiorillo\thanksref{addr2, addr3}
\and
F.~Galizzi\thanksref{addr5,addr6}
\and
I.~Gil-Botella\thanksref{addr1}
\and
C.~Gotti\thanksref{addr5}
\and
G.~Grauso\thanksref{addr3}
\and
J.~Jablonski\thanksref{addr9}
\and
A.A.~Machado\thanksref{addr7}
\and
S.~Manthey Corchado\thanksref{e1,addr1}
\and
J.~Mart\'in--Albo\thanksref{addr8}
\and
G.~Matteucci\thanksref{addr2,addr3}
\and
L.~Meazza\thanksref{addr5,addr6}
\and
A.P.~Mendoça\thanksref{addr7}
\and
A.~Minotti\thanksref{addr5,addr6}
\and
D.~Navas-Nicol\'{a}s\thanksref{addr1}
\and
L.~Pagliuso\thanksref{addr7}
\and
C.~Palomares\thanksref{addr1}
\and
L.~P\'{e}rez-Molina\thanksref{addr1}
\and
V.~Pimentel\thanksref{addr7}
\and
I.~L\'{o}pez~de~Rego\thanksref{addr1}
\and
Z.~Rautio\thanksref{addr9}
\and
J.~Romeo-Araujo\thanksref{addr1}
\and
D.~Rudik\thanksref{addr2,addr3}
\and
E.~Segreto\thanksref{addr7}
\and
Y.~Suvorov\thanksref{addr2,addr3}
\and
M.~Sturdivant\thanksref{addr9}
\and
F.~Terranova\thanksref{addr5}
\and
J.~Ureña\thanksref{addr8}
\and
A.~Verdugo~de~Osa\thanksref{addr1}
\and
D.~Warner\thanksref{addr9}
\and
R.~Wilson\thanksref{addr9}
\and
K.~Zhu\thanksref{addr9}
}

\thankstext{e1}{e-mail: sergio.manthey@proton.me}
\thankstext{e2}{e-mail: gabriel.botogoske@studenti.unipd.it}
\institute{
    \textbf{Centro de Investigaciones Energéticas Medioambientales y Tecnológicas (CIEMAT)}, 28040 Madrid (Spain) \label{addr1}
    \and
    \textbf{ Università degli Studi Federico II, Dipartimento di Fisica}, 80126 Napoli (Italy) \label{addr2}
    \and
    \textbf{ Istituto Nazionale di Fisica Nucleare, Sezione di Napoli}, 80126 Napoli (Italy) \label{addr3}
    \and
    \textbf{ University of Padova}, 35121 Padova (Italy) \label{addr4}
    \and
    \textbf{ Istituto Nazionale di Fisica Nucleare (INFN), Sezione di Milano-Bicocca}, 20126 Milano (Italy) \label{addr5}
    \and
    \textbf{ University of Milano Bicocca}, 20126 Milano (Italy) \label{addr6}
    \and
    \textbf{ University of Campinas (Unicamp)}, 13083-852 Campinas (Brazil) \label{addr7}
    \and
    \textbf{ Instituto de Física Corpuscular (IFIC), CSIC–Universitat de València}, Paterna 46980 (Spain) \label{addr8}
    \and
    \textbf{ Colorado State University}, Colorado 80523 (USA) \label{addr9}
}


\date{Received: date / Accepted: date}

\maketitle

\begin{abstract}
  The DUNE experiment will implement a photon detection system composed of \acl{XA} (\acs{XA}) devices. These trap incoming VUV photons by internal reflection in a wavelength shifter light guide to be collected onto silicon photomultiplier arrays, sensitive to visible light. In the baseline design, dichroic filters are used to prevent photons from escaping. The configuration proposed for \acs{DUNE}'s \acl{VD} (\acs{VD}) module has been characterised in liquid argon for the first time using dedicated cryogenic setups developed at \acs{CIEMAT} and \acs{INFN} Naples. Additionally, several alternative configurations, based on the design optimisation studies of an R\&D campaign, have been evaluated. The results show an efficiency of up to \SI{4.5(4)}{\percent} at \SI{4.5}{\volt} overvoltage, representing a significant improvement over previous \acs{XA} implementations. Most notably, configurations without dichroic filters show an improvement of up to \SI{18}{\percent}, attributed to transmittance losses in the dichroic filters.



\end{abstract}


\flushbottom
\section{Introduction\label{sec:INTRODUCTION}}
The \Ac{DUNE} is a next--generation dual--site experiment~\cite{DUNE:2020lwj} that aims to measure the neutrino oscillation parameters with unprecedented precision~\cite{2020arXiv200203005A}, perform \ac{BSM}~\cite{DUNE:2020fgq} searches as well as detect astrophysical neutrinos from the Sun and core-collapse supernovae within our galaxy~\cite{DUNE:2020zfm} by instrumenting four \SI{17}{\kilo\tonne} \acp{LArTPC} \ac{FD} modules. The first \ac{FD} module will adopt the \ac{VD} technology, in which ionisation charges are drifted perpendicularly from a central cathode to two anode planes. This generates two fiducial volumes (top and bottom) of approximately \SI{7}{\meter} in drift distance. 

The \ac{PDS} of \ac{DUNE}'s \ac{FD}-\ac{VD} module consists of light collector devices (\qtyproduct[product-units = power]{60x60}{\centi\meter}), called \ac{XA}~\cite{Machado_2018}, which are placed in the cathode structure (320 \acp{XA}) and at the inner sides of the cryostat walls (352 \acp{XA}). The \acp{XA} on the cathode must be sensitive to light produced in both drift regions, requiring that both surfaces be active, i.e., \ac{DS}. On the other hand, the \ac{XA} modules mounted on the wall only have one sensitive side, or \ac{SS}. 

\Ac{LAr} is an ideal target for neutrino detection because interacting particles generate ionisation electrons and \ac{VUV} scintillation light. The particle's energy deposits generate singlet and triplet states of Ar dimers ($\text{Ar}_{2}^{\ast}$) that de-excite emitting \ac{VUV} photons (\SI{127}{\nano\metre}) with characteristic times $\tau_\text{fast} = \SI{7.1}{\nano\second}$ and $\tau_\text{slow} = \SI{1.66}{\micro\second}$~\cite{Hitachi1983}. The prompt-light signal provides the trigger and interaction time for non-beam events, such as supernovae and solar neutrinos, thereby reconstructing the drift coordinates of those interactions. To evaluate the performance of the \ac{PDS}, the \ac{LY}, defined as the total \ac{PE} count detected onto the \ac{PDS} per energy deposit, is the main figure of merit to be considered. To meet the scope of \ac{DUNE}'s physics program, an average \ac{LY} of \SI{20}{\photoelectron\per\mega\electronvolt} is required in the \ac{FD}-\ac{VD} module, which can can be achieved with a \ac{PDE} of the \ac{XA} device \SI{>3}{\percent}~\cite{DUNE:2023nqi}.

This work presents the measurement of the absolute \ac{PDE} and optimisation of the \ac{DUNE} \ac{FD}-\ac{VD} \ac{XA} device in a cryogenic \ac{LAr} environment, performed by two research institutions: \ac{CIEMAT} in Madrid and \ac{INFN} in Naples. Each institution has developed its own measurement to ensure the reliability of the results. Section~\ref{sec:INTRODUCTION} describes the implementation of the \ac{XA} concept to \ac{DUNE}'s \ac{VD}-\ac{FD} module. The methodology is explained in Section~\ref{sec:METHODOLOGY}. In Section~\ref{sec:CIEMAT_SETUP}, the cryogenic setup implemented at \ac{CIEMAT} used to measure different \ac{XA} configurations is presented. Section~\ref{sec:ANALYSIS} shows the basic calibration and analysis steps. Section~\ref{sec:RESULTS} provides a comprehensive description of the \ac{PDE} measurements, and the \ac{XA} optimisation is analysed by comparing all \ac{XA} configurations measured at \ac{CIEMAT}. In the same Section, a cross-check of the \ac{CIEMAT} results' \ac{PDE} measurement performed in Naples is presented. Finally, Section~\ref{sec:CONCLUSION} concludes with this work's main observations aiming to improve the \ac{XA} design. The Appendices~\ref{sec:SIMULATION}, \ref{sec:CORRECTION}, and~\ref{sec:NAPLES_SETUP} provide complementary information on Naples laboratory setup, as well as required Geant4 simulations and analysis corrections, respectively.

\subsection{\acs{XA} Concept for \acs{DUNE}\label{sec:XA_CONCEPT}}
The \ac{XA} is a compact, modular and scalable photon detector that can be adapted to different configurations and applications~\cite{DUNE:2020txw}. It is designed to operate at cryogenic temperatures and can detect \ac{VUV} scintillation light with low noise. Its goal is to maximise the photon collection of the \ac{PDS} while minimising the amount of expensive \acp{SiPM} required to achieve the desired \ac{PDE}.

\begin{figure}[ht!]
    \centering
    \includegraphics[trim={7cm 1cm 7cm 5cm},clip,width=\linewidth]{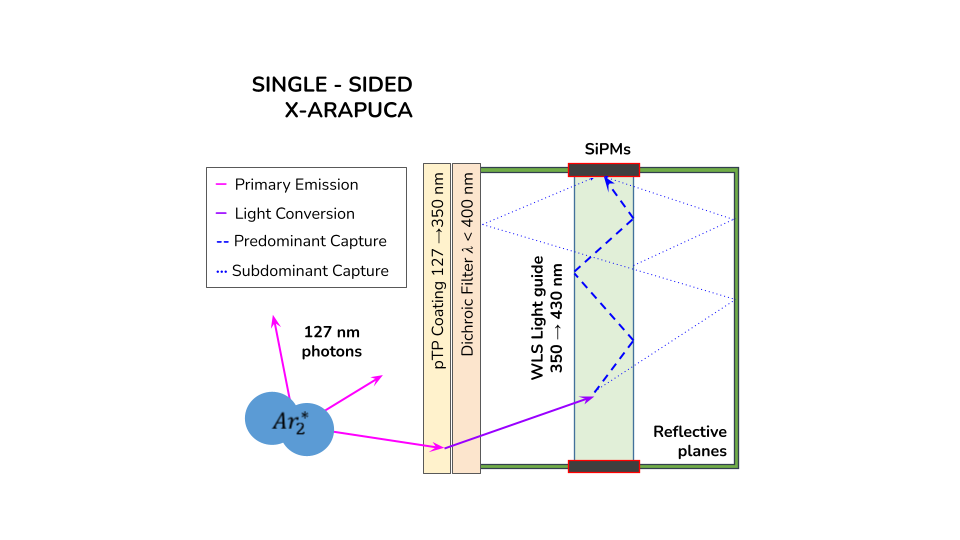}
    \caption{Schematic of the \ac{XA} working principle. Not to scale.}
    \label{fig:xarapuca_concept}
\end{figure}

The \ac{XA} entrance window is covered by a substrate coated on the external side with a \ac{pTP} layer (\SI{\sim500}{\micro\gram\per\square\centi\metre}) to convert the incident \SI{127}{\nano\meter} scintillation light into isotropically re-emitted \SI{350}{\nano\meter} photons with a high efficiency~\cite{NIJEGORODOV2000783}. The re-emitted light reaches the acrylic (polymethyl methacrylate) \ac{WLS--LG} where it is further downshifted and isotropically re-emitted to the visible range (\SI{430}{\nano\meter}) thanks to the doping with a chromophore (see Figure~\ref{fig:xarapuca_concept}). 

The \ac{WLS--LG} re-emitted light can be trapped by total internal reflection (critical angle $\theta\sim56^{\circ}$) and reach the surrounding \ac{SiPM} arrays or escape and be reflected back into the \ac{WLS--LG} by a thin--film \ac{DF} coating, whose cutoff should be above \SI{400}{\nano\meter}. The non-active surfaces of the \ac{XA} are covered with a highly reflective \href{https://multimedia.3m.com/mws/media/982449O/3mtm-specular-film-df2000ma-technical-data-sheet.pdf}{Specular Film} (VIKUITI\textsuperscript{T}) to reduce the absorption of the escaping photons and redirect them to the \ac{SiPM} for enhanced collection probability. The \ac{XA} technology has demonstrated its superior \ac{PDE} compared to dip-coated shift light guides~\cite{DUNE:2020cqd}. 

\begin{figure}[ht!]
    \centering
    \includegraphics[width=\linewidth]{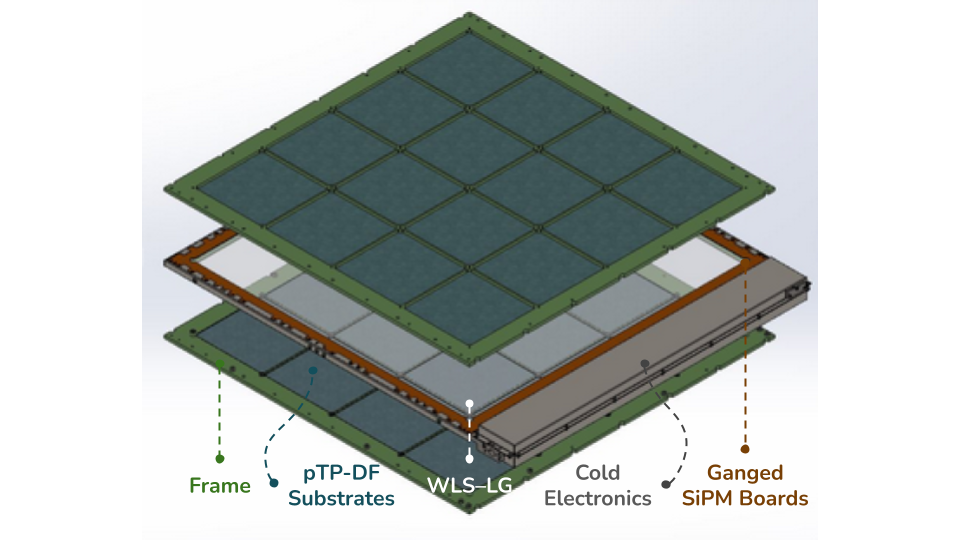}
    \caption{Scheme of the \ac{XA}-\ac{DS} implementation for \ac{DUNE} \ac{FD}-\ac{VD}  module. Displayed are the substrates (top and bottom) and the \ac{WLS--LG} layer in the middle. Substrates are surrounded by the \ac{XA} mounting frame and the \ac{WLS--LG} by the \ac{SiPM} strips.}
    \label{fig:xa_reference}
\end{figure}

Figure~\ref{fig:xa_reference} shows the \ac{XA} implementation for the \ac{VD}-\ac{FD} module, whose surface (\qtyproduct[product-units = power]{60x60}{\centi\meter}) is covered by 16 \ac{pTP}-coated \acp{DF} (\qtyproduct[product-units = power]{14.6x14.6}{\centi\meter}) arranged over a G10 grid frame. In each \ac{XA}, \num{160} \acp{SiPM} are mounted on 8 flex-boards and positioned around the \ac{WLS--LG}. These \acp{SiPM} (\href{https://www.fbk.eu/en/}{FBK} Triple--Trench~\cite{Acerbi_2026}) have an active area of \qtyproduct[product-units = power]{6x6}{\milli\meter}. The \ac{XA} mechanical frame consists of brackets holding the different components in place. The \ac{XA} baseline design includes \acp{DF} from \href{https://www.zaot.com/en/}{ZAOT} manufacturer~\cite{ZAOT} and the \SI{3.8}{\milli\meter} thick \ac{WLS--LG} with a chromophore concentration of \SI{80}{\milli\gram\per\kilo\gram} from \href{https://www.glasstopower.com/}{G2P}~\cite{G2P}.

The mounted \acp{SiPM} are ganged in groups of five and deliver output data across two channels connected to the readout cold-amplifier. The signals are routed to the front-end electronics by custom-designed signal leading-boards~\cite{Gallice:2021tad}. This cold stage is composed of a main board with two cold transimpedance amplifiers, which are responsible for the active ganging of the \acp{SiPM}. Each input reads 80 \acp{SiPM} (half of the \ac{XA}). The cold amplifier is held inside a metal casing mechanically attached to one side of the \ac{XA}.

\subsection{\acs{XA} Simulation\label{sec:XA_SIMULATION}}
To estimate the device's response to the different optical components and inform about the specific \ac{XA} configuration to be tested, a \ac{MC} simulation, described in detail in a forthcoming publication, of the whole \ac{XA} has been developed to compute the device's \ac{PDE}. This simulation implements the measured optical properties of all active components. 

\begin{figure}[ht!]
    \centering
    {\includegraphics[width=\linewidth]{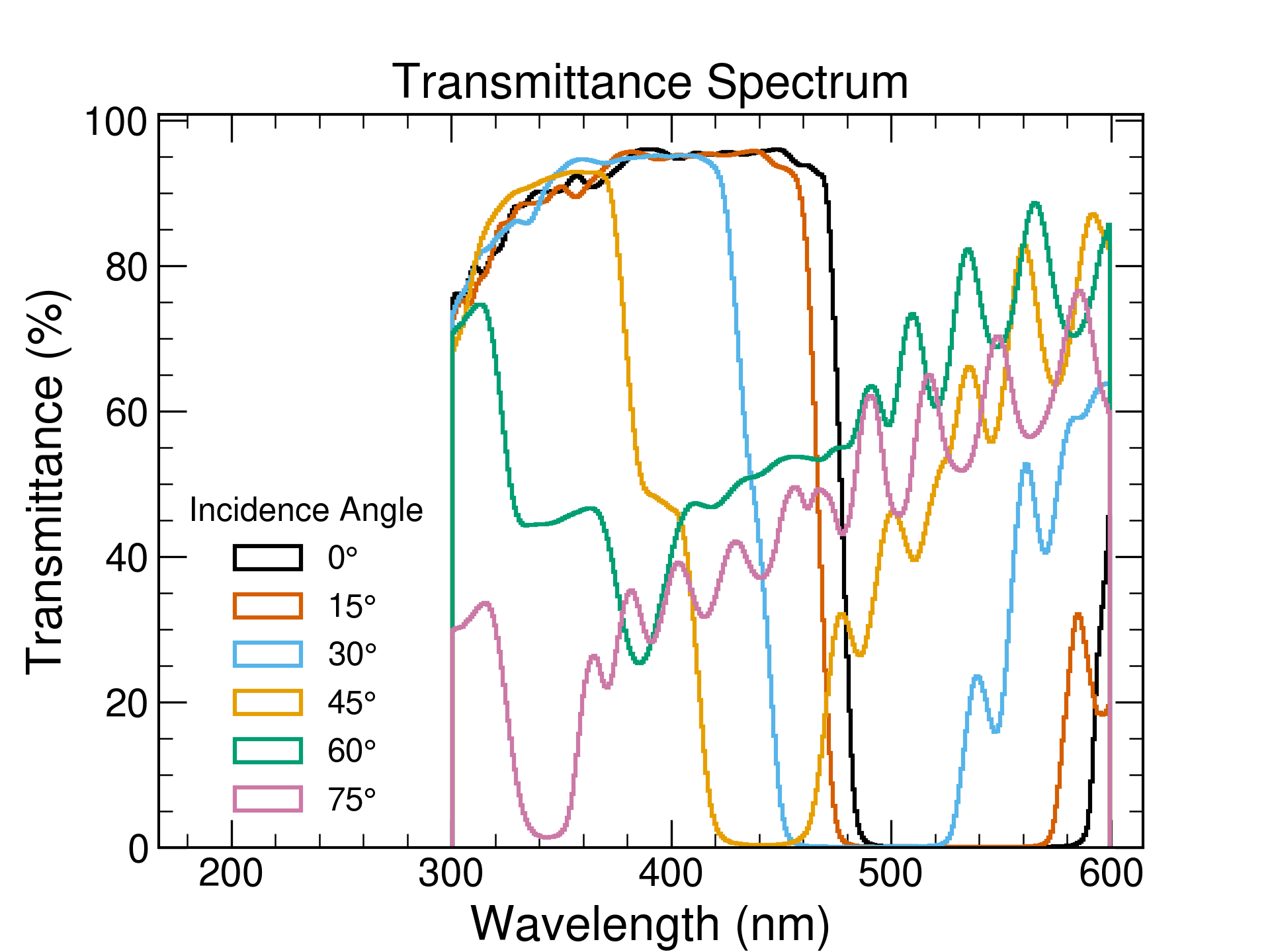}}
    \caption{\ac{DF} transmission curves from \href{https://www.zaot.com/en/}{ZAOT} manufacturer measured in demineralised water for different angles of incidence~\cite{Cattadori_2024}. The corresponding curves in \ac{LAr} can be converted from these using Snell's law (e.g. \ang{45} curve corresponds to \ang{49} in \ac{LAr}).} 
    \label{fig:DF_transmittance}
\end{figure}

Figure~\ref{fig:DF_transmittance} shows the transmittance of the implemented \ac{DF} as a function of the incidence angle. These \acp{DF} must operate in a medium (\ac{LAr}) with a high refractive index ($\rm n$\num{\sim1.3}) and for light over a broad range of incidence angles. As a consequence, the measured transmittance curves exhibit a markedly non-ideal behaviour in two respects. First, the cut-off value shifts with the angle of incidence, compromising the target value of 400 nm, which is only maintained at \ang{45}. Second, the non-zero transmittance above the cut-off wavelength allows photons to escape, thereby reducing the filter’s confinement efficiency.

The simulation predicts that removing the dichroic coating increases the \ac{PDE} by approximately \qtyrange{10}{15}{\percent}. This effect arises from the inherent angular dependence of the \ac{DF} transmission curves combined with the broad angular distribution of the interacting photons. Specifically, the isotropic re-emission of photons by \ac{pTP} results in a fraction being reflected by the \ac{DF}, preventing their entry into the \ac{XA}. Similarly, photons isotropically re-emitted by the \ac{WLS--LG} exhibit a wide angular spread, causing a fraction to be transmitted through the \ac{DF} and thus escape the \ac{XA}. Even if the use of a \ac{DF} enhanced the trapping of outgoing photons compared to a clear substrate, this improvement does not seem to compensate for the loss of incoming photons reflected by the \ac{DF} in the first place. These studies have motivated the measurement of \acs{XA} without \acp{DF}. 

Another factor influencing the \ac{XA} \ac{PDE} is the \ac{WLS--LG} chromophore concentration. The trapping efficiency of \ac{pTP} emitted photons increases with the chromophore concentration, but at the same time, the attenuation length of the converted light is reduced, resulting in a counteractive behaviour that needs to be optimised. Three different concentrations of chromophore: 24, 40, and 80 \si{\milli\gram\per\kilo\gram} have been tested with the simulation for different \ac{WLS--LG} thicknesses. The initial results suggest that a lower dye concentration with an increased optical path (thicker \ac{WLS--LG}) can improve the \ac{XA} performance. 

Furthermore, the simulation predicts a compatible performance for double and single-sided configurations. 



\subsection{\acs{XA} Configurations\label{sec:CIEMAT_CONFIGURATIONS}}
The predicted improvement in \ac{PDE} for a configuration without \ac{DF} has been assessed with the production of \ac{pTP}-coated substrates without the addition of dichroic layers. These have been mounted and tested for both \ac{SS} and \ac{DS}  configurations of the \ac{XA}. Additionally, two \ac{WLS--LG} designs have been produced by \href{https://www.glasstopower.com/}{G2P}, one being the baseline choice, and the second has been produced with lower chromophore concentration and a thicker bar. Both are presented in Table~\ref{tab:wls_configurations}.

\begin{table}[htp]
    \centering
    \resizebox{\columnwidth}{!}
    {
        \begin{tabular}{lcc}
            \multicolumn{3}{c}{\acl{WLS--LG} Summary} \\
            \hline
            \ac{WLS--LG}    & Length x Height x Width           & Chromophore                        \\ \hline \hline
            \textbf{A} & \qtyproduct[product-units = power]{605x605x3.8}{\milli\metre} & \SI{80}{\milli\gram\per\kilo\gram} \\
            \textbf{B} & \qtyproduct[product-units = power]{607x605x5.5}{\milli\metre} & \SI{24}{\milli\gram\per\kilo\gram} \\ \hline
        \end{tabular}
    }
    \caption{Tested \ac{WLS--LG} configurations. Showing the baseline configuration (A) and the alternative model (B). The chromophore concentration is given in \si{\milli\gram\per\kilo\gram} of the acrylic substrate.}
    \label{tab:wls_configurations}
\end{table}

\begin{figure}[ht!]
    \centering
    {\includegraphics[width=0.8\linewidth]{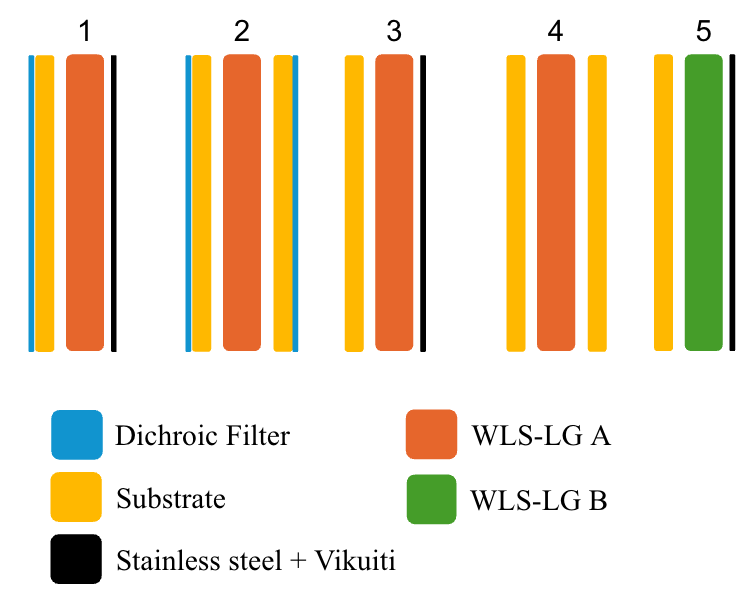}}
    \caption{Schematic representation of tested \ac{XA} configurations at CIEMAT site.\label{fig:xa_combinations}}
\end{figure}

Table~\ref{tab:xa_combinations} shows the list of all tested \ac{XA} configurations, and a diagram with the different elements is shown in Figure~\ref{fig:xa_combinations}. It is important to mention that the \ac{SS} \ac{XA} baseline design with \ac{DF} has been measured in the first and last campaigns to verify the reproducibility of the results across two different assemblies. 

\begin{table}[ht!]
	\centering
	\resizebox{\columnwidth}{!}
    {
    \begin{tabular}{llccr}
        \multicolumn{5}{c}{\acl{XA} Configurations} \\
        \hline
          & Configuration   & Filter & WLS--LG & Type          \\\hline \hline
        1 & DF-XA-SS           & Yes    & A       & Single--Sided \\
        2 & DF-XA-DS        & Yes    & A       & Double--Sided \\
        3 & noDF-XA-SS         & No     & A       & Single--Sided \\
        4 & noDF-XA-DS      & No     & A       & Double--Sided \\ 
        5 & noDF-XA-SS\_24mg   & No     & B       & Single--Sided \\\hline
    \end{tabular}
	}
	\caption{Tested \ac{XA} configurations at CIEMAT site.\label{tab:xa_combinations}}
\end{table}

\section{Methodology\label{sec:METHODOLOGY}}
 
The \ac{PDE} is defined as the number of \acp{PE} measured by the photosensor device (the \ac{XA} in our case) with respect to the true number of incident photons arriving at its surface (PDE$_\text{XA} = \text{PE}_\text{XA}/\gamma_\text{true}$).

We distinguish between two computation strategies: reference and simulation. These differ in the approach of measuring the true number of incident photons. Experimentally, the two methods have been realised by constructing two cryogenic setups that use $^{241}$Am source to produce \ac{VUV} scintillation photons in \ac{LAr}. This ensures the \ac{PDE} is assessed to reproduce the operating conditions of the \ac{DUNE} experiment.

In both \ac{PDE} computation methods the number of true collected \acp{PE} ($\text{PE}_\text{XA}$) is determined from the number of detected \acp{PE} in the \ac{XA} ($\text{PE}'_\text{XA}$) and a factor ($f^\text{XT}$) to correct for the known probability of a secondary pulse, \ac{XT}, in the \ac{XA} \acp{SiPM} as follows:

\begin{ceqn}
    \begin{equation}
    \text{PE}_\text{XA}  = \dfrac{Q}{Q^{\text{SPE}}}\cdot  \left(1-P^\text{XT}_\text{XA}\right) \equiv \text{PE}'_\text{XA} \cdot f^\text{XT}_\text{XA}\,,
    \end{equation}
\end{ceqn}

\noindent where $Q$ corresponds to the mean charge of the scintillation photons collected by the \ac{XA}; $Q^{\text{SPE}}$ is the corresponding $\text{SPE}$ charge value and $P^\text{XT}$ is the \ac{XT} probability.

In the reference method, the number of incident photons in the \ac{XA} is obtained from the \acp{PE} measured by a reference \ac{VUV}-sensitive \ac{SiPM} with known efficiency (PDE$_\text{SiPM}$). The true number of incident photons can be expressed in terms of the \acp{PE} detected by the reference \acp{SiPM} ($\gamma_\text{true} = \text{PE}_\text{SiPM}/\text{PDE}_\text{SiPM}$). A geometric factor ($f^\text{geo}$) accounts for the different active surfaces of the sensors and their relative positions. This method has the advantage of being less sensitive to global deficits in the predicted light level that could be introduced by impurities in \ac{LAr}. The disadvantage is that the uncertainty in the reference \ac{SiPM}'s efficiency is directly propagated to the \ac{XA} \ac{PDE} measurement (as will be shown in Section~\ref{sec:RESULTS}). Crucially, these uncertainties are cancelled when comparing different \ac{XA} configurations. Therefore, the corresponding \ac{PDE} computation can be formulated as follows:

\begin{ceqn}
    \begin{equation}
        \text{PDE}^\text{reference}_\text{XA} = \frac{\text{PE}'_\text{XA}\cdot f^\text{XT}_\text{XA}}{\text{PE}'_\text{SiPM}\cdot f^{XT}_\text{SiPM}}\cdot \text{PDE}_\text{SiPM} \cdot f^\text{geo}, \label{eq:ciemat_pde}
    \end{equation}
\end{ceqn}

\noindent for which $f^\text{geo}$, the ratio of the geometrical acceptances per unit area, has been evaluated with a simulation (see Appendix~\ref{sec:CIEMAT_SIMULATION}). 

In the case of the simulation method, the true number of incident photons is computed from the theoretical light-yield generated by an alpha particle and a \ac{MC} simulation that takes into account the full geometry of the \ac{XA} and is detailed in Appendix~\ref{sec:NAPLES_SIMULATION}. This method is highly sensitive to variations in the total scintillation light yield, which can be affected by \ac{LAr} impurities that do not influence the slow light component and therefore cannot be monitored. A purification system and an analytical purity correction are needed to ensure that real setup measurements agree with the light production expected from the simulation. To evaluate the \ac{PDE}, the following calculation is performed:

\begin{ceqn}
    \begin{equation}
        \text{PDE}^\text{simulation}_\text{XA}=\frac{\text{PE}'_\text{XA} \cdot f^\text{XT}_\text{XA}}{\text{MC}^\text{photons}} \cdot \dfrac{1}{f^\text{bending}} \cdot \dfrac{1}{f^\text{purity}},
        \label{eq:naples_pde}
    \end{equation}
\end{ceqn}

\noindent where $f^\text{purity}$ is the purity correction of the liquid argon (see Section~\ref{sec:NAPLES_PDE}), and $f^\text{bending}$ is an additional correction required to compensate for the bending of the \ac{XA} light guide (see Appendix~\ref{sec:CORRECTION}).

\section{Experimental Setup\label{sec:CIEMAT_SETUP}} 



\begin{figure}[htp]
	\centering
    \includegraphics[trim={2.5cm 0 2.5cm 0},clip,width=\linewidth]{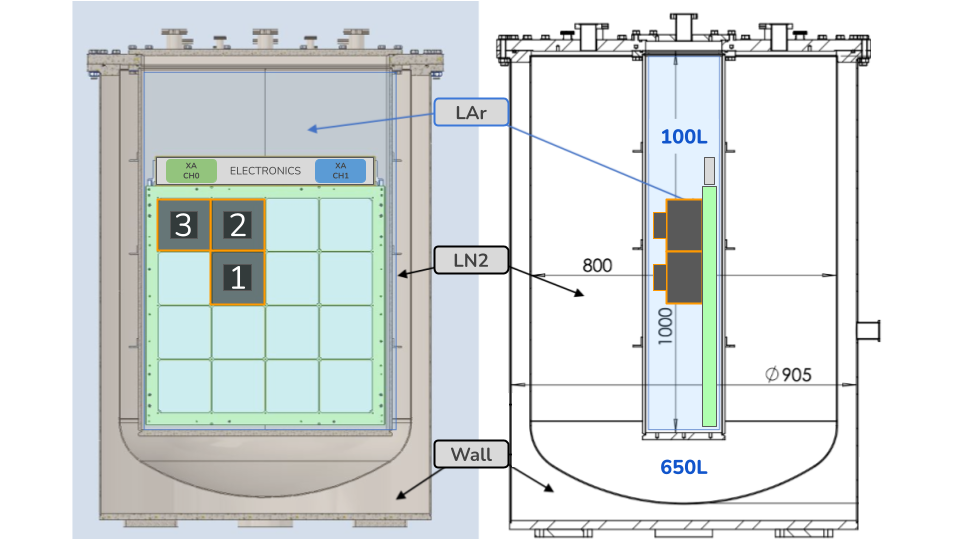}
	\caption{Setup scheme of the cryogenic vessels available at \ac{CIEMAT} for the \ac{XA}'s absolute efficiency measurement.}
	\label{fig:vessel_graphic}
\end{figure}

The experimental setup (see Figure~\ref{fig:vessel_graphic}) consists of a \SI{\sim650}{\liter} cylindrical vessel with an internal squared cassette (\SI{\sim100}{\liter}) that hosts the \ac{XA} where grade 6.0 \ac{GAr} can be liquefied. The external vessel is filled with \ac{LN$_2$} to serve as a cold reservoir for the inner volume. Before filling, successive vacuum cycles (\SI{<2e-5}{\milli\bar}) are performed to avoid outgassing of the setup components, which might worsen the \ac{LAr} purity. The \ac{GAr} liquefaction happens due to thermal contact with the LN$_2$ reservoir. The setup is designed to maintain constant operating conditions (pressure and temperature) during the liquefaction process. The cold reservoir is filled with LN$_2$ from an external tank and maintained at \SI{2.7}{\bar} to prevent the argon from freezing during liquefaction. The process is monitored by PT100 temperature sensors that provide live temperature readings of the \ac{LAr} level inside the vessel.

\begin{figure}[htp]
    \centering
    \includegraphics[trim={8cm 0cm 0cm 0cm},clip,width=\linewidth]{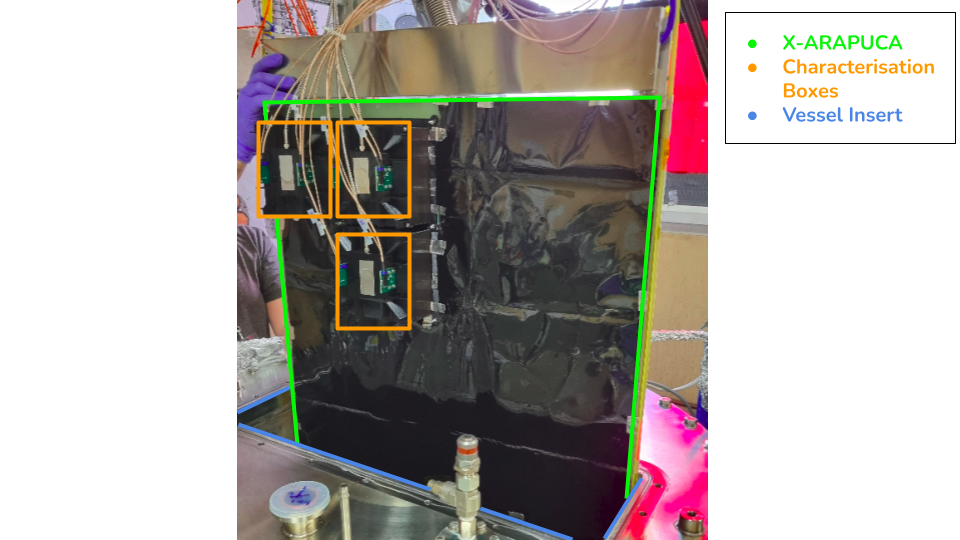}
    \caption{Insertion of an \ac{XA} inside the cryogenic vessel.}
    \label{fig:xa_insertion}
\end{figure}

\begin{figure}[htp]
	\centering
    \includegraphics[trim={3cm 0cm 3cm 0},clip,width=\linewidth]{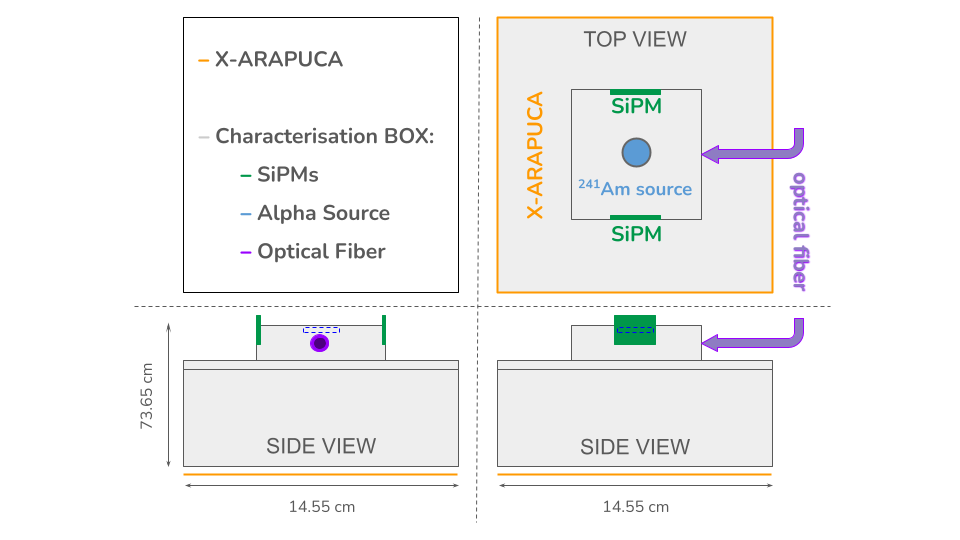}
	\caption{Characterisation box hosting the reference sensors together with the $^{241}$Am source and the inlet for the optical fibre used in calibration.}
	\label{fig:box_graphic}
\end{figure}

The \ac{XA} is introduced vertically (see Figure~\ref{fig:xa_insertion}) together with three characterisation boxes (one for each of the uniquely distinct substrate-window positions of the \ac{XA}). Each black box (\qtyproduct[product-units = power]{14.6x14.6x7.6}{\centi\meter}) hosts a $\sim$55 Bq radioactive $^{241}$Am source~\cite{CalvoAlamillo:2023olz} that emits alpha-particles with energies \SI{5.468}{\mega\electronvolt}, \SI{5.443}{\mega\electronvolt} and \SI{5.388}{\mega\electronvolt} and ratios of \SI{85.2}{\percent}, \SI{12.8}{\percent} and \SI{1.7}{\percent}, respectively. Together with the $^{241}$Am source, two reference cryogenic VUV4 \acp{SiPM} (S13370--6075CN~\cite{Alvarez-Garrote:2024byb}) from Hamamatsu (HPK~\cite{HPK}) are placed inside each box (as represented in Figure~\ref{fig:box_graphic}). These \acp{SiPM} are selected because of their high sensitivity to \ac{VUV} light and stable performance at \ac{CT}, making it possible to directly detect the scintillation light produced in \ac{LAr} by the alpha particles emitted from the source. The scintillation photons produced by the source also reach the \ac{XA} since one of the sides of the characterisation box is open to the \ac{XA} window. To ensure that no other photons are detected, a black sheet covers the rest of the \ac{XA}. For calibration purposes, an optical fibre guides light from an external laser source (\SI{405}{\nano\meter}) that can be operated by a waveform generator, providing a precise and sharp pulse of light and the appropriate trigger to the \ac{DAQ} (\href{https://www.caen.it/products/dt5725/}{CAEN} DT5725~\cite{CAEN}) simultaneously. 

The sensors' dimensions and positions in the box with respect to the alpha source have been chosen to ensure appropriate illumination of the active surfaces with a Geant4 simulation (see Appendix~\ref{sec:CIEMAT_SIMULATION}).

Regarding the reference \acs{SiPM}, studies have shown that the efficiency of these \ac{VUV} sensors at CT decreases with respect to the \ac{PDE} measured by the manufacturer at \ac{RT}~\cite{Pershing:2022eka}. In addition, the PDE depends on the incident angle of the photons~\cite{nEXO:2019jhg}. The absolute \ac{PDE} of these \ac{VUV} reference \acp{SiPM} has been measured for \SI{127}{\nano\meter} at \SI{87}{\kelvin} and \SI{4}{\volt} \ac{OV} in a previous work~\cite{Alvarez-Garrote:2024byb}, and corrected for the average incident angle of the photons in the current setup (\ang{75}~\ref{sec:CIEMAT_SIMULATION}). The main calibration results of these reference sensors are listed in Table~\ref{tab:sipm_calibration}. The gain and the signal-to-noise ratio ($\rm S/N$) are computed from the charge of the \ac{SPE} spectrum. The cross-talk probability ($\rm P^{XT}$) is estimated from the same calibration data using the Vinogradov method~\cite{vinogradov2009probability}. The reference \acp{SiPM} have always been operated at the same \ac{OV}, ensuring that the \ac{PDE} is computed under the same conditions. 


\begin{table}[htp]
    \centering
    \resizebox{\columnwidth}{!}{  
        \begin{tabular}{ccccc}
            \multicolumn{5}{c}{\href{https://www.hamamatsu.com}{HPK} VUV4 (reference \acp{SiPM}) }  \\\hline
            \ac{OV} (V) & Gain ($\times 10^{6}$) & $\rm S/N$ & $\rm P^\text{XT}_\text{SiPM}$ (\unit{\percent}) & PDE$_\text{SiPM}$ (\unit{\percent}) \\ \hline \hline
            4.0 & \num{8.0(3)}  & \num{6.5(2.1)} & \num{25(3)} & \num{12.1(1.1)} \\ \hline
        \end{tabular}
    }
    \caption{Reference \acp{SiPM} calibration results at \SI{87}{\kelvin} and \SI{127}{\nano\meter} light.}
    \label{tab:sipm_calibration}
\end{table}

\begin{figure}[htp]
	\centering
	\includegraphics[trim={3cm 1cm 3.75cm 1.4cm},clip, width=1\linewidth]{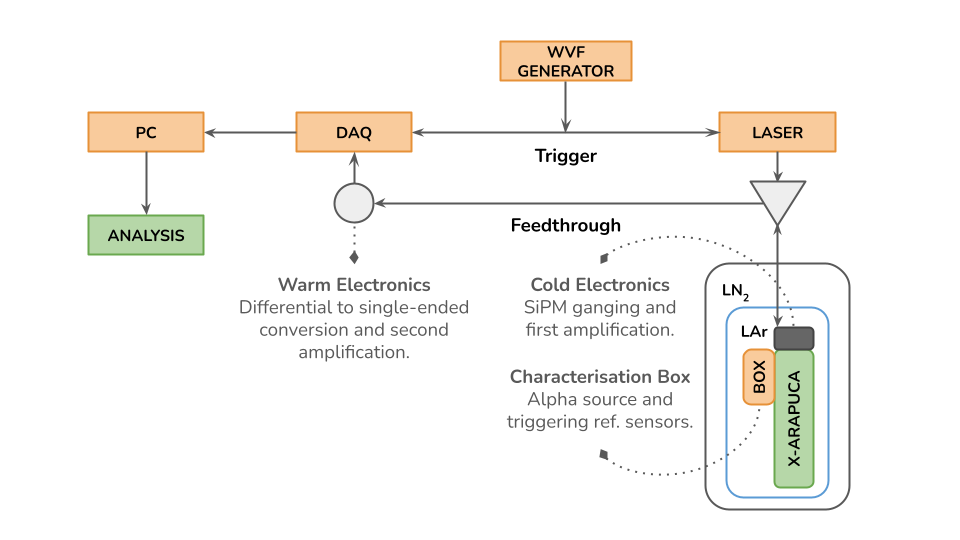}
	\caption{Schematic representation of the data flow during data acquisition.}
	\label{fig:DAQ}
\end{figure}

Each \ac{XA} configuration is tested in data campaigns lasting three days to ensure data stability and repeatability. The data transmission chain and \ac{DAQ}, involving the warm-amplification stage and the read-out system, are schematically represented in Figure~\ref{fig:DAQ}.

The setup implemented for the simulation method at INFN Naples also deploys the \ac{XA} in \ac{LAr} with an alpha source. The details of this setup are described in Appendix \ref{sec:NAPLES_SETUP}. 
\section{Data Analysis} \label{sec:ANALYSIS}
In the following, we present the common analysis steps of the data acquired at both \ac{CIEMAT} and \ac{INFN} setups. The \ac{XA} has been operated with voltage values of \SI{30.5}{\volt}, \SI{31.5}{\volt}, \SI{33}{\volt} and \SI{34}{\volt}, corresponding to \SI{3.5}{\volt}, \SI{4.5}{\volt}, \SI{6}{\volt} and \SI{7}{\volt} overvoltage, respectively\footnote{For reference, \SI{3.5}{\volt}, \SI{4.5}{\volt} and \SI{7}{\volt} overvoltage correspond to \SI{40}{\percent}, \SI{45}{\percent} and \SI{50}{\percent} efficiency of the mounted \acp{SiPM} at room temperature.}.

A standard set of data taken in a single campaign is shown in Table~\ref{tab:runs}. The calibration and alpha source data are used in the \ac{PDE} computation, and the muon and noise data monitor the purity and stability of the setup. 

\begin{table}[htp]
    \centering
    \resizebox{\columnwidth}{!}{%
        \begin{tabular}{lccc}
            \multicolumn{4}{c}{Data Type Summary} \\
            \hline
            Type             & Source     & Trigger  & Signal        \\ \hline \hline
            Calibration      & Laser      & External & Pulsed        \\
            $\alpha$--source & $^{241}$Am & \ac{DAQ} & Scintillation \\
            Muons            & Cosmic     & \ac{DAQ} & Scintillation \\
            Noise            & None       & Random   & None          \\ \hline
        \end{tabular}%
    }
    \caption{Main data types and their characteristics.}
    \label{tab:runs}
\end{table}

\subsection{\acs{XA} Calibration\label{sec:COMMON_PROCEDURES}}
\begin{figure}[ht]
    \centering
    \includegraphics[width=\linewidth]{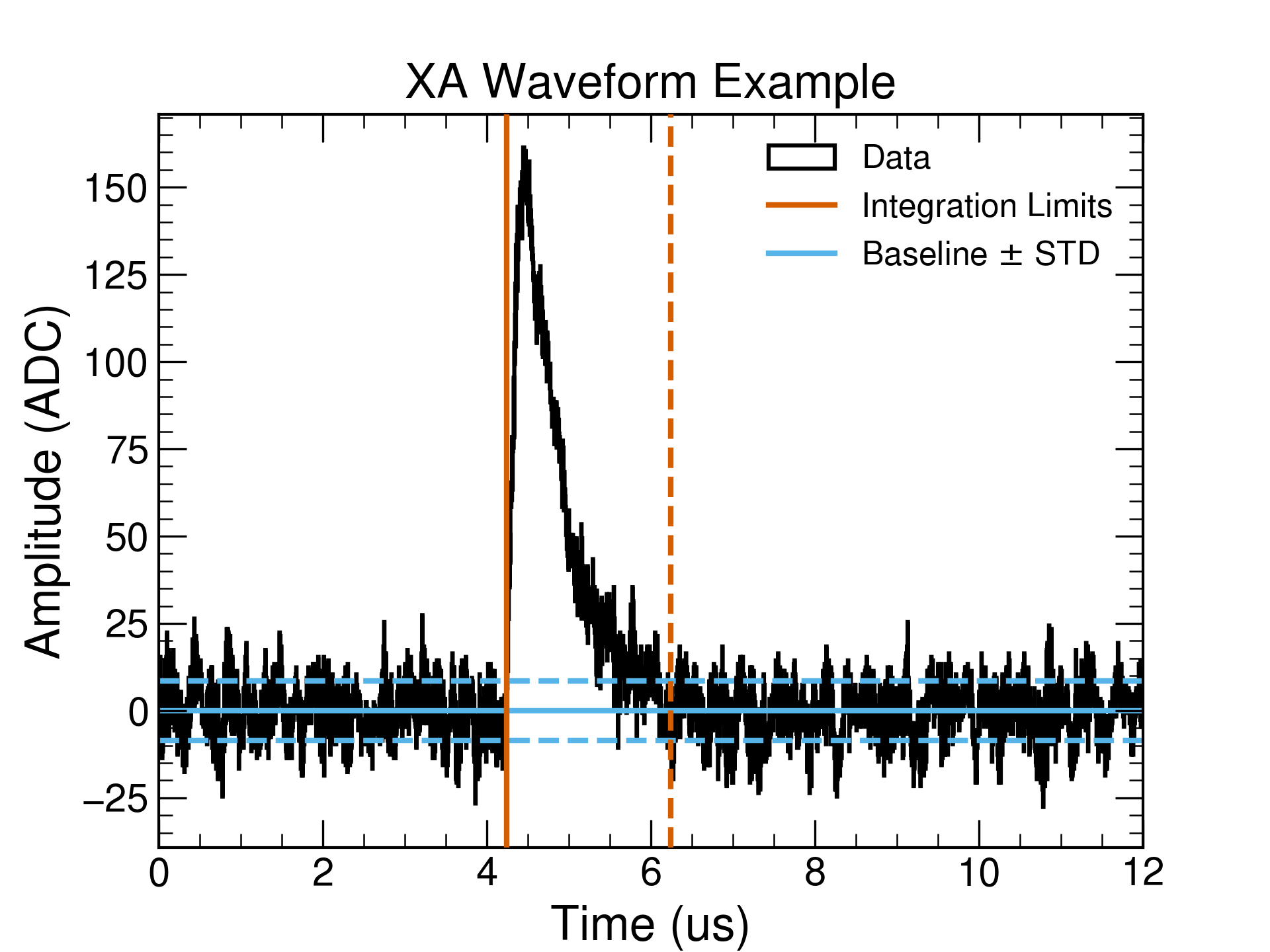}
    \caption{\ac{XA} signal example from calibration data at \SI{4.5}{\volt} overvoltage.}
    \label{fig:XA_waveform}
\end{figure}

The first step in the analysis is the \acs{XA} gain calibration with data taken by illuminating the \acs{XA} with a low-intensity pulsed laser source. The triggering scheme is provided by a pulsed generator that simultaneously feeds the laser and the \ac{DAQ}. The result is a set of aligned waveforms with a defined time correlation. Figure~\ref{fig:XA_waveform} shows an example waveform for calibration data of the \ac{XA}. 

\begin{figure}[ht]
    \centering
    \includegraphics[width=\linewidth]{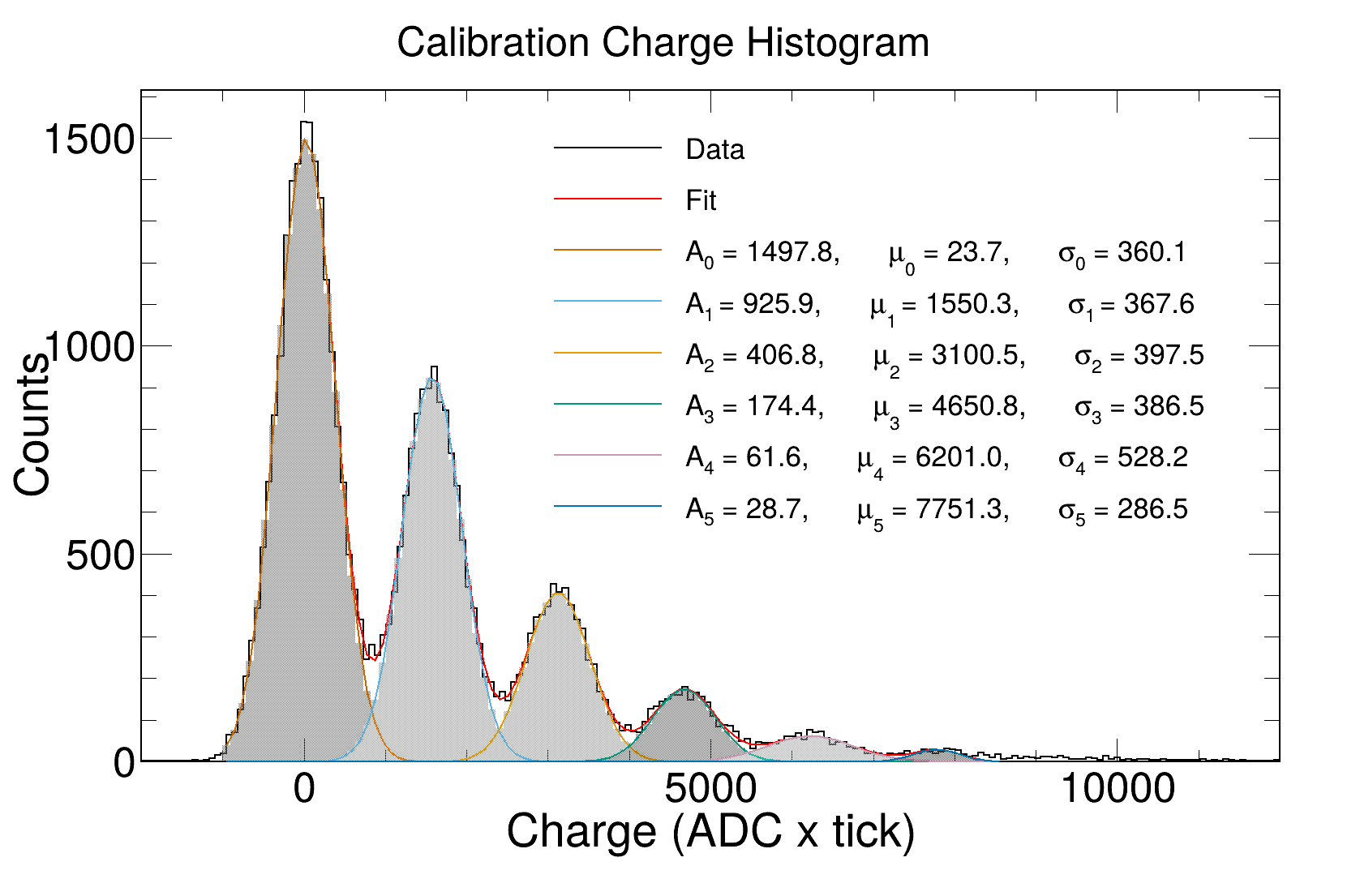}
    \caption{Charge histogram for the \ac{XA} at an overvoltage of 4.5V}
    \label{fig:spe}
\end{figure}

\begin{figure}[htp]
    \centering
    \includegraphics[width=\linewidth]{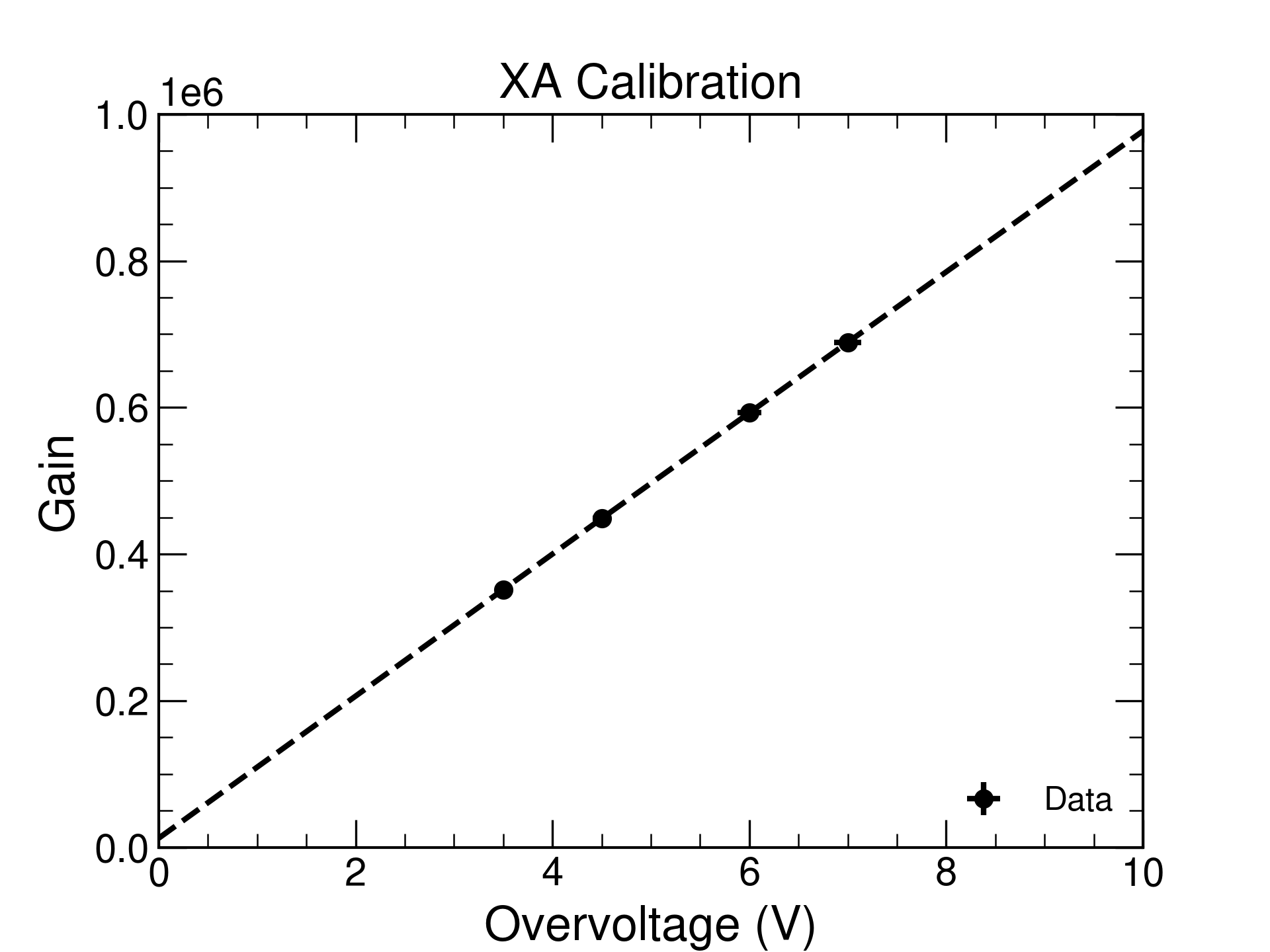}
    \caption{Gain versus bias voltage, for the baseline \ac{XA} configuration. The dashed line shows the function computed for the best-fit parameters.}
    \label{fig:XA_gain}
\end{figure}

The charge integral is calculated on a fixed time window from the signal's rise time. To determine the gain, the mean charge of the \ac{SPE} peak of the charge histogram is divided by the electron's charge and electronic amplification. The charge histograms are computed and fitted individually with Gaussian curves for each \ac{XA} readout channel (see Figure~\ref{fig:spe}). The evaluated gain of the \ac{XA} as a function of the overvoltage is shown in Figure~\ref{fig:XA_gain}. The signal-to-noise ratio is determined as:

\begin{ceqn}
    \begin{equation}
        \text{SNR} = \frac{\mu_{1}}{\sqrt{\sigma_0^2+\sigma_1^2}},
        \label{eq:SNR}
    \end{equation}
\end{ceqn}

\noindent where $\mu_{1}$ is the mean of the first $Q^\text{SPE}$ distribution and $\sigma_\text{i}$ are the corresponding Gaussian sigmas for the pedestals and first \ac{SPE} peak.

Considering the same charge histogram, the cross-talk is calculated using the Vinogradov method from the signal density (normalised charge of the Gaussian fits) over the number of \acp{PE}. To obtain the real distribution of \acp{PE} (subtracting \ac{XT} effects), it is necessary to consider the factor $f^\text{XT} = 1- P^\text{XT}$. The final set of characterisation data extracted from calibration data is summarised in Table~\ref{tab:xa_calibration}.

\begin{table}[ht!]
    \resizebox{.49\textwidth}{!}
    {
        \begin{tabular}{lccc}
            \multicolumn{4}{c}{\ac{XA} Calibration (DF--XA)}                            \\\hline
            \ac{OV} (V) & Gain ($\times 10^{5}$) & $\rm S/N$     & $\rm P^{XT}$ (\unit{\percent}) \\ \hline \hline
            3.5    & $3.52\pm 0.03$        & 3.1 $\pm$ 0.8 & 12.8 $\pm$ 0.3    \\
            4.5    & $4.48\pm 0.04$        & 3.7 $\pm$ 0.9 & 18.4 $\pm$ 0.4    \\
            6.0    & $5.75\pm 0.04$        & $4.1 \pm 0.2$ & 29.0 $\pm$ 0.2    \\
            7.0    & $6.89\pm 0.06$        & 5.4 $\pm$ 1.3 & 32.6 $\pm$ 0.7    \\ \hline
        \end{tabular}
    }
    \caption{Average \ac{XA} calibration results from combined analysis of both \ac{XA} channels.}%
    \label{tab:xa_calibration}
\end{table}

\subsection{Scintillation Charge Analysis} \label{sec:SCINTILLATION}

The charge spectrum from the alpha particles' scintillation light is obtained to estimate the device's \ac{PDE}. For this measurement, the light of the $^{241}$Am source is triggered and evaluated.

In the reference method, a characterisation box ensures the fixed position of the $^{241}$Am source with respect to the photosensitive area of the \ac{XA} and the two reference sensors. The reference \acp{SiPM} provide a coincidence trigger that excludes all noise from the analysis sample. Due to the reduced size of the box, any cosmic interaction is also quite constrained and, if present, easily removable with an amplitude condition. The charge of the scintillation light is obtained by integrating the waveforms of the two \ac{XA} channels individually and combining their average values extracted from each Gaussian-fitted distribution. 

 Due to the finite size of the alpha source (non-point-like) and the \ac{SiPM} positions within the characterisation box, the charge spectra of these present an asymmetric shape. Unlike the case of the \ac{XA} response, their respective charges need to be first combined to recover a symmetric spectrum that can be fitted to a Gaussian. 
 
 The number of detected PEs ($\text{PE}'$) is then computed from the mean charge of the scintillation photons over the \ac{SPE} charge.



In Naples' setup, the position of the $^{241}$Am source holder needs to be calibrated to ensure its stability (see Appendix~\ref{sec:NAPLES_SETUP}). The source height has been adjusted to different values above the \ac{XA} (\qtylist{5;10;15}{\centi\meter}). At each height, measurements were taken at four different positions above the \ac{XA} surface (see Appendix~\ref{sec:NAPLES_SETUP}). These different light measurements are compared against the simulation results of the setup (see Appendix~\ref{sec:NAPLES_SIMULATION}). The alpha signal is obtained by triggering on one channel, and the charge is reconstructed by independently integrating the waveforms from each \ac{XA} channel over a \SI{6.6}{\micro\second} window and then combining the means. 

\begin{figure}[ht]%
    \centering
    \includegraphics[width=\linewidth]{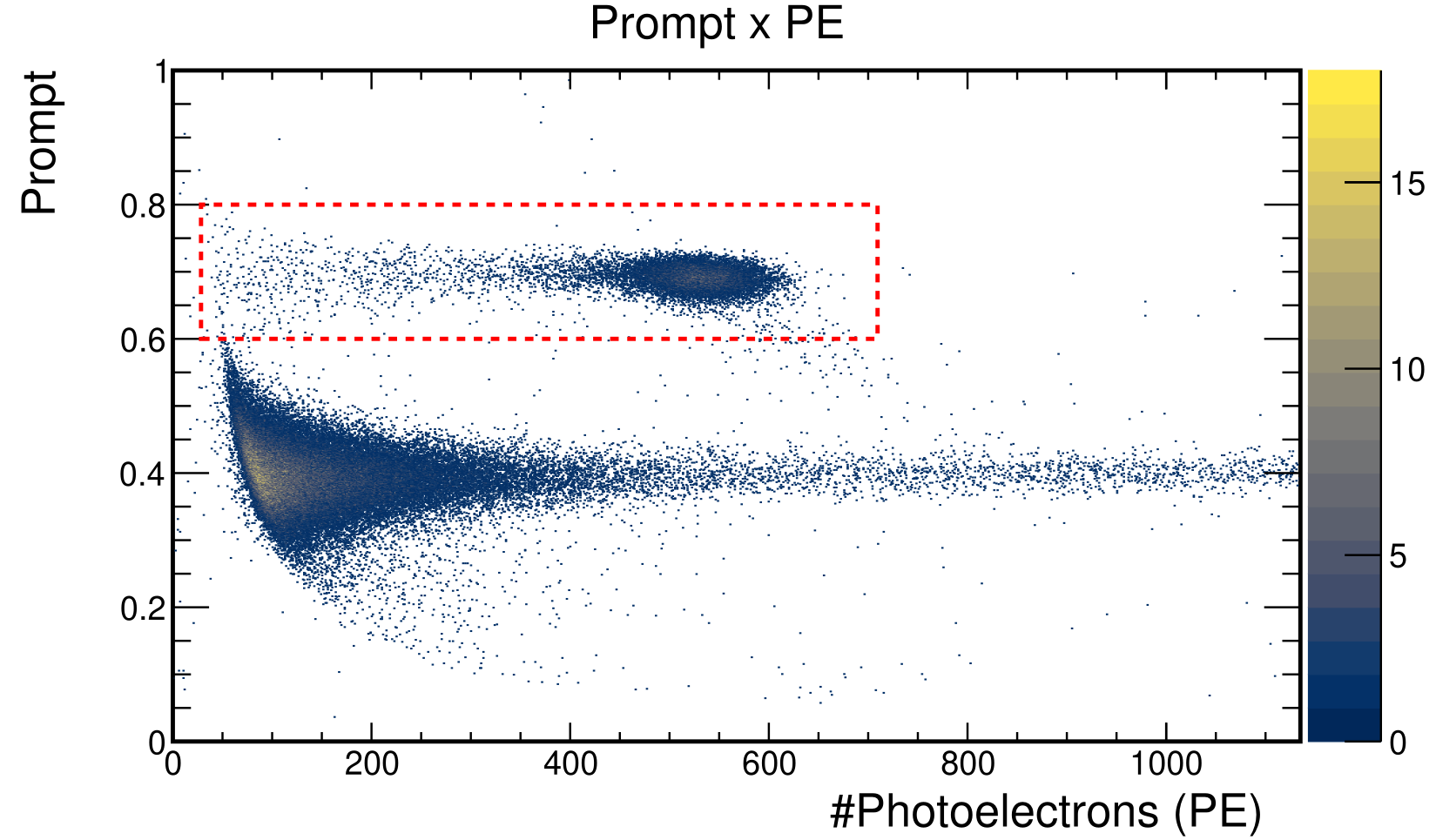} %
    \caption{Prompt distribution as a function of the total charge. In red is the selected region due to the scintillation of alpha particles.\label{fig:arapuca_fprompt}}
\end{figure}

By calculating the prompt ratio, defined as the ratio of the fast integral over the total integral, it is possible to select a region where argon nuclear recoils occur more frequently than electron recoils, allowing for the separation of alpha particles from cosmic muons. This is because alpha particles are more likely to interact via nuclear recoil than muons. The prompt ratio is calculated according to Equation~\ref{eq:fprompt}:

\begin{ceqn}
    \begin{equation}
        \text{Prompt} = \frac{\int_{t_{0}}^{t'}v(t)dt}{\int_{t_{0}}^{T}v(t)dt}\,,
        \label{eq:fprompt}
    \end{equation}
\end{ceqn}

\noindent where $v(t)$ is the measured signal, $t_{0}$ is the start time of the waveform, and $T$ is the waveform end time (\SI{6.6}{\micro\second}). The parameter $t'$ represents the upper limit for the fast integral (\SI{0.92}{\micro\second}). 
An example of the excellent alpha/muon separation that this method provides is shown in Figure \ref{fig:arapuca_fprompt}.

\begin{figure}[ht]%
    \centering
    \includegraphics[width=1\linewidth]{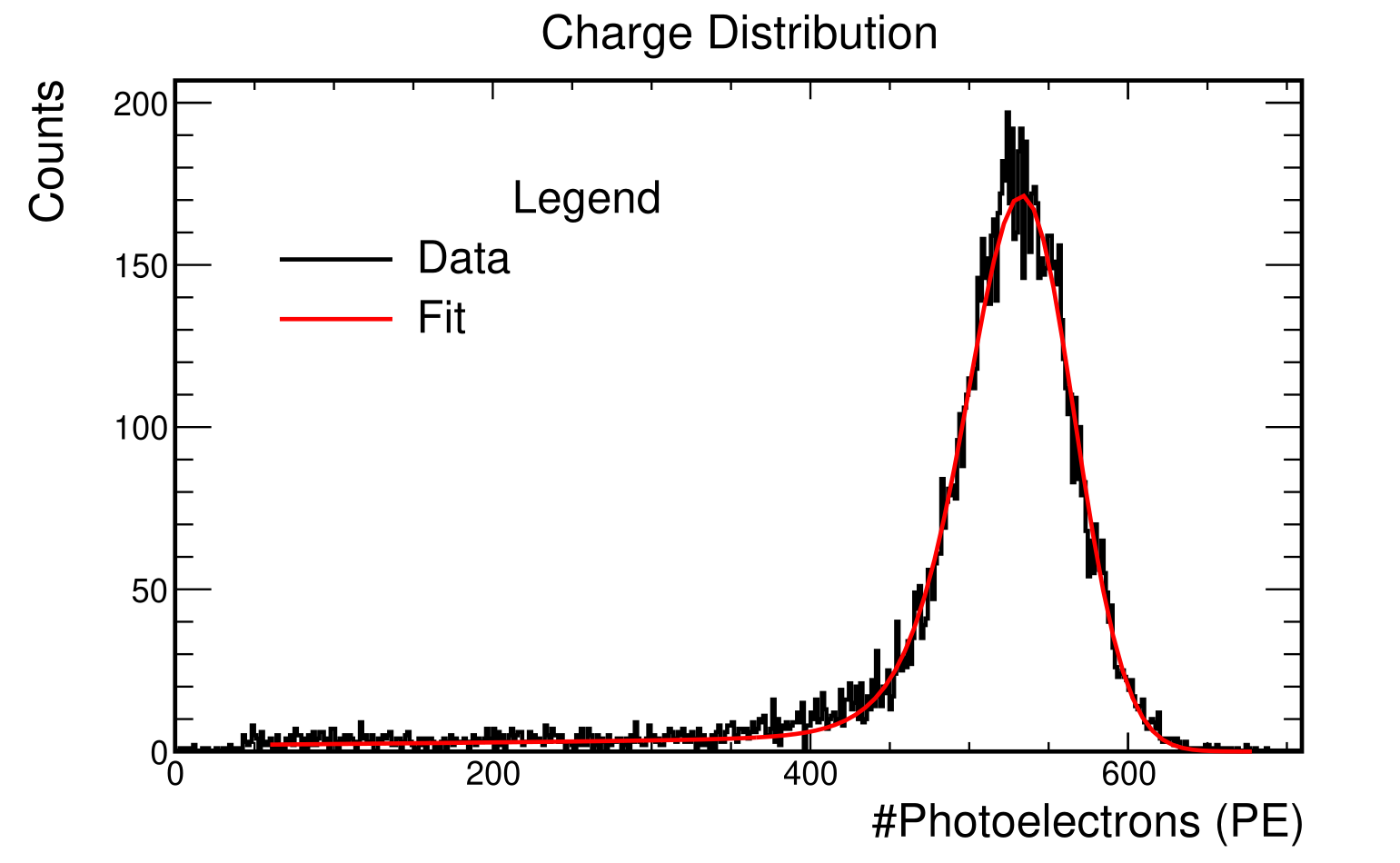} %
    \caption{Alpha's charge spectrum (\ac{XA} overvoltage of \SI{7}{\volt} and alpha source at position 2 (see Appendix~\ref{sec:NAPLES_SETUP}) and \SI{15}{\centi\meter} above the \ac{XA}), from channel 0.}
    \label{fig:integral_spectrum_alpha}%
\end{figure}

The signal induced by alpha particles was selected on events with $\text{Prompt} > 0.6$. After the selection, the charge spectrum results in a Gaussian with a long tail extending toward lower energies due to edge effects of the source holder, as shown in Figure~\ref{fig:integral_spectrum_alpha}. To obtain the peak's central value, the distribution of the number of photoelectrons obtained was fitted with Equation~\ref{eq:alpha_fit}, which is the sum of two exponential functions multiplied by a complementary error function.

\begin{align}
    y(Q) = \sum_{i=1}^{2} A_i \cdot 
    &\exp\left( \frac{Q - \mu}{\tau_i} + \frac{\sigma_i^2}{2\tau_i^2} \right) \notag \\
    &\cdot \erfc\left( \frac{1}{\sqrt{2}} 
    \left( \frac{Q - \mu}{\sigma_i} + \frac{\sigma_i}{\tau_i} \right) \right)
    \label{eq:alpha_fit}
\end{align}

The $A_1$ and $A_2$ values are the renormalisation factors of both functions; $\mu$ is the central position of the peak, $\sigma_1$ and $\sigma_2$ are the widths of both Gaussians, and $\tau_1$ and $\tau_2$ are the values responsible for the long tail.
\section{Results\label{sec:RESULTS}}
There are three main \ac{XA} design choices that have been evaluated in this work: The performance of the cryostat wall (\ac{SS}) and cathode (\ac{DS}) \acp{XA}, the effect of the \ac{DF} on the \ac{PDE}, and the dependence of the \ac{WLS--LG} dye concentration on the collection efficiency. In the following, we discuss the measurement results of the \ac{PDE} for the baseline design, \ac{SS} and \ac{DS}, and its variation across the different configurations tested using the reference method. The stability and repeatability of the measurements have also been studied. The simulation method measurement is also shown as a cross-check of the reference method results.

\subsection{Measurement of the PDE with the Reference Method \label{sec:RESULTS_CONFIGURATION}}
\begin{figure}[ht!]%
    \centering
    \includegraphics[width=1\linewidth]{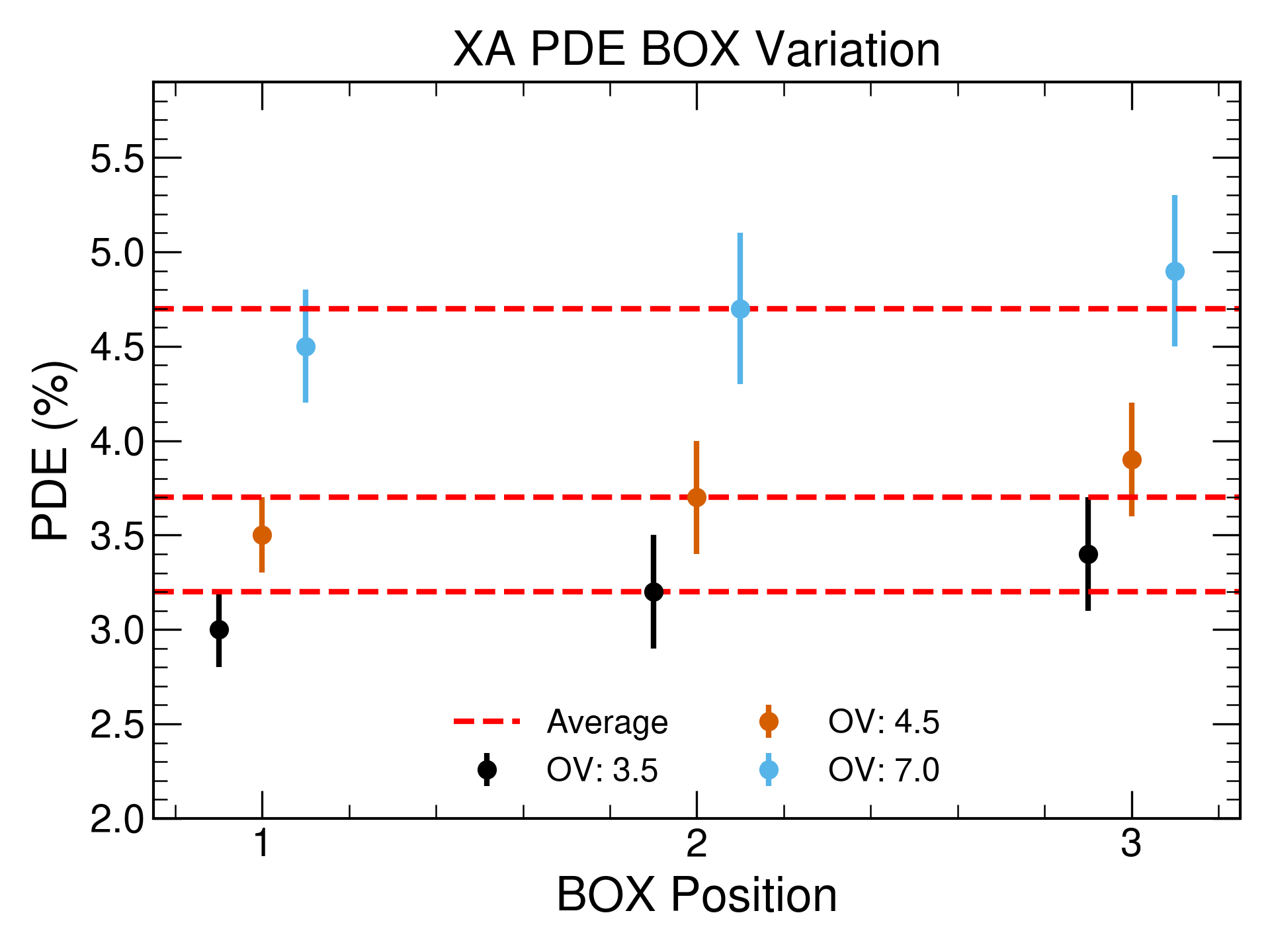}
    \caption{Baseline \ac{XA} \ac{PDE} from the reference method. Showing the \ac{PDE} as a function of characterisation box position. The dashed line represents the average position--weighted value of the \ac{PDE}.}
    \label{fig:box_pde}%
\end{figure}

We evaluate the measured \ac{PDE} values of the \ac{XA} designs by averaging over the three characterisation box results taking into account their respective \ac{XA} surface contributions (see Figure\ref{fig:box_pde}). Table~\ref{tab:results} shows the \ac{PDE} results for all the tested \ac{XA} configurations, obtained with the reference method. The errors are about \SI{9}{\percent} except in the case of the \ac{DF}-\ac{XA}-\ac{SS} configuration (\SI{\sim 7}{\percent}), which has been measured twice. The main source of error is the uncertainty on the reference \acp{SiPM} efficiency, PDE$_\text{SiPM}$ (\SI{8.7}{\percent}), obtained from previous measurements. It can be seen that the values of the \ac{PDE} for the cryostat wall (\ac{SS}) and cathode (\ac{DS}) \ac{XA} are well above the required value (\SI{3}\percent). 

\begin{table}[ht!]
\centering
    \begin{tabular}{rlccc} 
        \multicolumn{5}{c}{VD--XA PDE Results (\unit\percent)}       \\\hline
        & Config          & \ac{OV} \SI{3.5}{\volt} & \ac{OV} \SI{4.5}{\volt} & \ac{OV} \SI{7}{\volt}\\ \hline \hline
        1 & DF-XA-SS      & 3.2 $\pm$ 0.2      & 3.7 $\pm$ 0.3      & 4.7 $\pm$ 0.3   \\ 
        2 & DF-XA-DS      & 3.5 $\pm$ 0.3      & 4.0 $\pm$ 0.4      & 5.0 $\pm$ 0.5   \\
        3 & noDF-XA-SS    & 3.9 $\pm$ 0.4      & 4.5 $\pm$ 0.4      & 5.8 $\pm$ 0.6   \\
        4 & noDF-XA-DS    & 3.8 $\pm$ 0.4      & 4.5 $\pm$ 0.4      & 5.6 $\pm$ 0.6   \\
        5 & noDF-XA\_24mg & 3.6 $\pm$ 0.4      & 4.3 $\pm$ 0.4      & 5.5 $\pm$ 0.6   \\\hline
    \end{tabular}
    \caption{Results for the \ac{XA}-\ac{VD} absolute \ac{PDE} for all tested configurations and \ac{OV} values. \label{tab:results}}
\end{table}




To better illustrate the differences between the various configurations, Table~\ref{tab:results_comparison} shows the relative variations in \ac{PDE} for the three distinct overvoltages and their average. Notice that the overvoltage is not expected to have any impact on the relative variation of the \ac{PDE}. Relative errors are smaller than the ones of the absolute results due to the cancellation of the PDE$_\text{\ac{SiPM}}$ term in the PDE$_\text{\ac{XA}}$ result.


\begin{table}[ht!]
\centering
    \resizebox{\columnwidth}{!}{
        \begin{tabular}{lcccc} 
        \multicolumn{5}{c}{Relative PDE Improvement}\\\hline
                      & \multicolumn{2}{c}{(no\ac{DF}-\ac{DF})/\ac{DF}}    &\multicolumn{2}{c}{(\ac{DS}-\ac{SS})/\ac{SS}}\\
            \ac{OV} (\unit{\volt})   & \ac{SS}       & \ac{DS}       & \ac{DF}      & no\ac{DF}     \\\hline\hline
            \num{3.5} & \SI{18(3)}{\percent} & \SI{8(3)}{\percent}  & \SI{7(3)}{\percent} & \SI{-2(3)}{\percent} \\ 
            \num{4.5} & \SI{17(3)}{\percent} & \SI{12(3)}{\percent} & \SI{3(3)}{\percent} & \SI{-2(3)}{\percent} \\
            \num{7.0} & \SI{20(3)}{\percent} & \SI{12(3)}{\percent} & \SI{3(3)}{\percent} & \SI{-4(3)}{\percent} \\\hline
             Mean     & \SI{18(3)}{\percent}  & \SI{11(3)}{\percent}  & \SI{4(3)}{\percent}  & \SI{-3(3)}{\percent}  \\\hline
        \end{tabular}
    }
    \caption{Relative improvements for the \ac{XA}'s absolute efficiency for different configurations and \ac{OV} values. \label{tab:results_comparison}}
\end{table}

The most important results of this comparison concern the difference between no-\ac{DF} and \ac{DF} devices (columns two and three in Table~\ref{tab:results_comparison}). One concludes that the \acp{XA} without \ac{DF} presents a noticeable increase in \ac{PDE} (\SI{18}{\percent} and \SI{11}{\percent}) with respect to the \ac{SS} and \ac{DS} counterparts. As anticipated in Section~\ref{sec:INTRODUCTION}, these results were expected because of the non-ideal behaviour of the implemented \acp{DF}. We have computed and confirmed this effect, strengthening the methodology and analytic tools developed for the optimisation and development procedures. 

Next, the \ac{PDE} for the \SI{24}{\milli\gram\per\kilo\gram} chromophore concentration and \SI{5.5}{\milli\meter} width \ac{WLS--LG} configuration shows no remarkable change with respect to the baseline design, pointing to the need for better implementation of the \ac{WLS--LG} in the simulation. Further investigation of the result shows a different behaviour in the measured \ac{PDE} depending on the position of the alpha source. The variation from box 1 to box 3 is up to \SI{13}{\percent} for the baseline \ac{WLS--LG} (see Figure~\ref{fig:box_pde}), while a flatter response (\SI{<5}{\percent}) is measured for the lower chromophore concentration. This can be attributed to the expected larger attenuation length. This is a desirable feature for the \ac{DUNE} \ac{PDS}, as it allows for a more uniform light collection efficiency across the detector's volume. 

The \ac{DS} and \ac{SS} configurations present compatible \acp{PDE} results (within errors), as expected from simulation. The small differences observed between the \ac{DF} and no\ac{DF} variations (\SI{4}{\percent} and \SI{-3}{\percent} relative improvement in \ac{PDE}) cannot be resolved with the current data. The fact that both configurations present similar results points to the preference for the detection of photons trapped through internal reflection in the \ac{WLS--LG}. In other words, efforts to collect photons that have escaped the \ac{WLS--LG} by exceeding the medium’s critical angle are not efficient enough to result in measurable improvements.

\par
The next step is to evaluate the stability and reproducibility of the measurements. This is crucial for ensuring that the \ac{PDE} results are consistent over time. The gain monitoring of the sensors, as well as a stable and consistent \ac{PDE} evaluation during all data runs, have been deemed necessary to consider a campaign successful. 


\begin{figure}[ht!]
    \centering
    \includegraphics[width=1\linewidth]{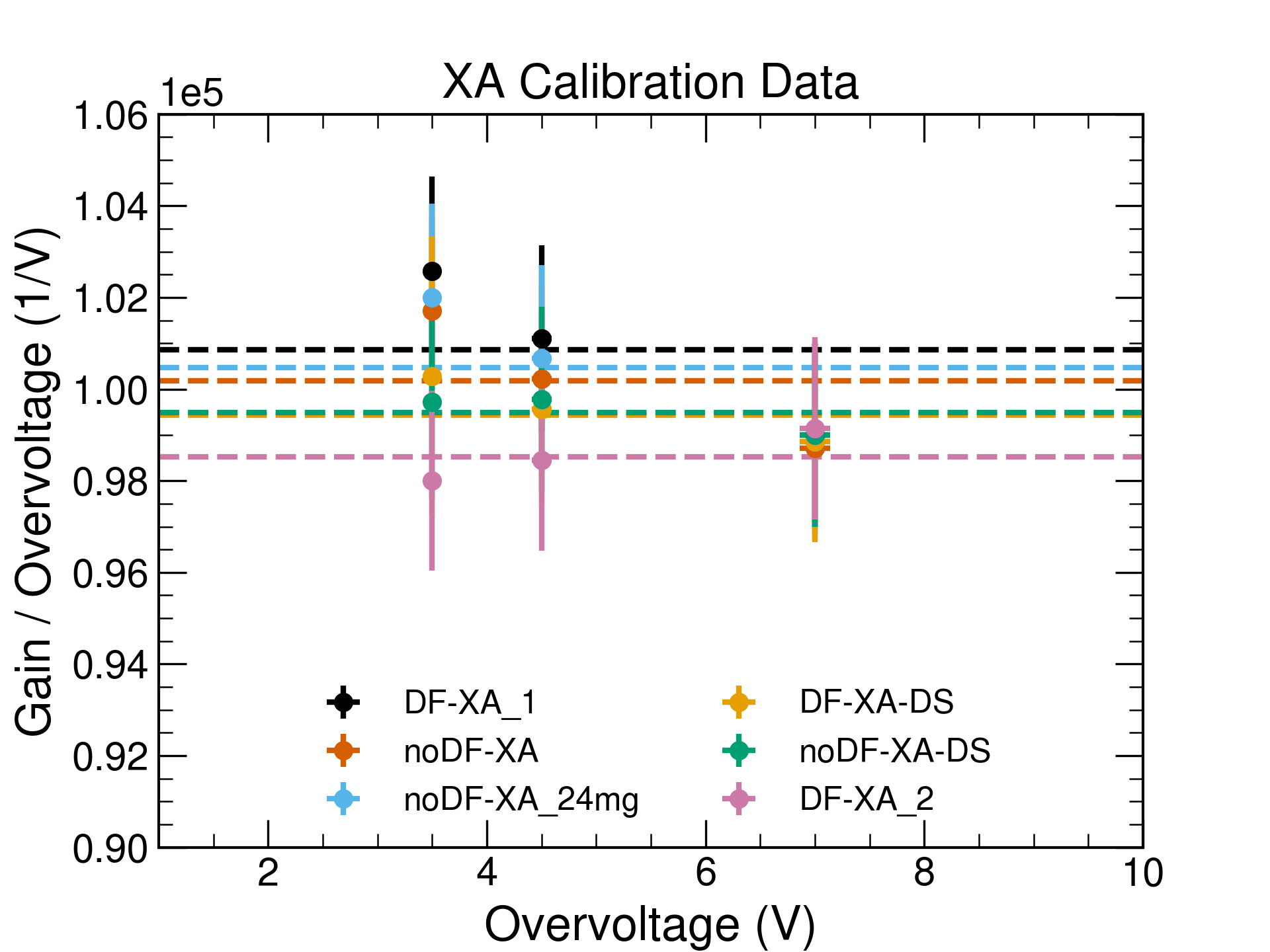}
    \caption{\ac{CIEMAT}'s gain/\ac{OV} slope for all tested \ac{XA} configurations. Dashed lines represent the weighted average.}
    \label{fig:gain_stability}%
\end{figure}

In the case of \ac{CIEMAT}'s results, the readout can be tested along with the different campaigns. The gain/\ac{OV} slope of the \ac{XA} calibration is shown in Figure~\ref{fig:gain_stability}. The results show a consistent gain/\ac{OV} slope across all configurations (\SI{\pm2}{\percent}), indicating that the experimental setup is operated at stable conditions and that the electronics of the \ac{XA} are not drifting over the \SI{\sim 1}{\year} period of data taking. 

\begin{figure}[htp]
	\centering
	\includegraphics[width=1\linewidth]{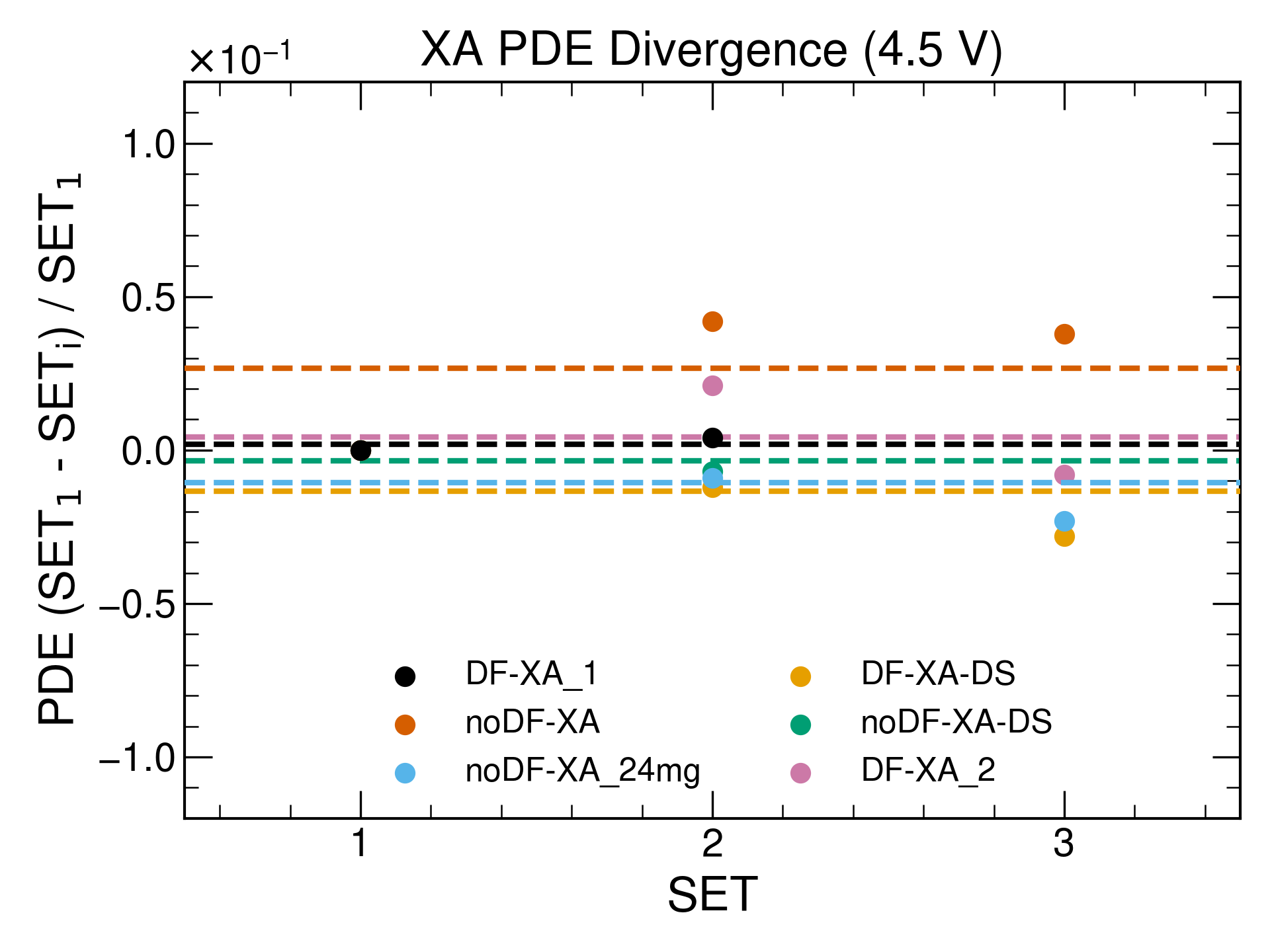}
	\caption{Relative \ac{PDE} deviation with respect to the first evaluated value for all \ac{XA} configurations at \SI{4.5}{\volt} overvoltage. The dashed lines represent the mean of the deviations.}
	\label{fig:PDE_xa}
\end{figure}

Three sets of runs for each \ac{XA} configuration have been acquired. Figure~\ref{fig:PDE_xa} presents a set of results (for \SI{4.5}{\volt} \ac{OV}) that show the relative divergence of the \ac{PDE} values for the second and third sets for all \ac{XA} configurations with respect to the first computed value. As can be observed, the relative \ac{PDE} variations fall within a \SI{3}{\percent} mean range that lies reasonably below the computed uncertainty of the \ac{XA} \ac{PDE} measurement (\SI{\sim9}{\percent}). 


\subsection{Measurement of the PDE with the Simulation Method\label{sec:NAPLES_PDE}}

In this section, we present the results obtained with the simulation method at the Naples setup. As explained in Section~\ref{sec:METHODOLOGY}, the true number of incident photons is determined from a \ac{MC} simulation considering the expected light-yield for the alpha source. Two correction factors should be taken into account, $f^\text{purity}$ and $f^\text{bending}$. The \ac{LAr} purity is a critical parameter, as the presence of contaminating impurities can affect the absolute light propagation~\cite{ref:n2_contamination}. 
This effect can be quantified by analysing the slow scintillation component of liquid argon ($\tau_{slow}$), as impurities quench its intrinsic decay time of \SI{1660(100)}{\nano\second}~\cite{Hitachi:1983zz}. 
To determine the slow component, the average waveform was calculated using signals from a Hamamatsu PMT R11065 inside the vessel, operated at \SI{1500}{\volt} and triggered by cosmic muon interactions in \ac{LAr}. The muons were selected based on the region with a prompt light ratio close to \num{0.4} (see Figure~\ref{fig:arapuca_fprompt}). 



\begin{figure}[ht]%
    \centering
    \includegraphics[width=1\linewidth]{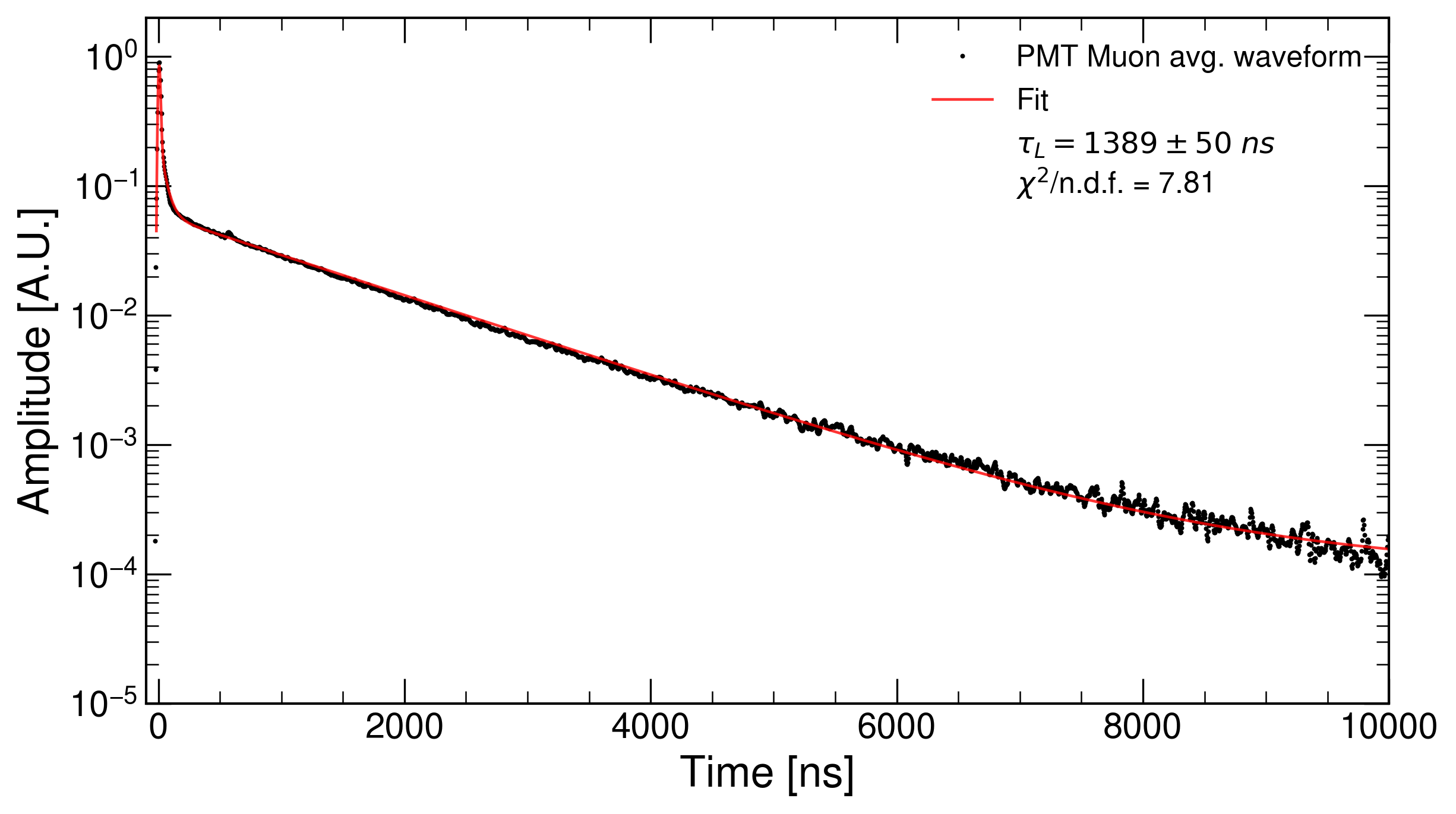} %
    \caption{Fitted \ac{LAr} slow component of the scintillation light captured by the \ac{PMT}.}%
    \label{fig:pmt}%
\end{figure}

The value of the slow component in the setup ($\tau'_{slow}$) was obtained by fitting the tail of the result of the convolution of the slow component with a Gaussian filter (simulating the electronics response) and with the \ac{pTP} response~\cite{ref:tpb_emissium}. The fitted average waveform is shown in Figure~\ref{fig:pmt} and the value obtained is \SI{1389(50)}{\nano\second}. The liquid argon purity correction factor is therefore estimated with the equation:

\begin{ceqn}
    \begin{equation}
        f^\text{purity} 
        = A_\text{fast} + A_\text{slow} \cdot\frac{\tau'_{slow}}{\tau_{slow}}\\
        =  0.94 \pm 0.06
    \label{eq:lar_purity}
    \end{equation}
\end{ceqn}

\noindent where the proportion between fast and slow components was calculated using the mean values obtained at \cite{ref:liquid_argon_1,ref:liquid_argon_2,ref:liquid_argon_3}. 

A bending of the \ac{WLS--LG} producing its misalignment with the \acp{SiPM} was observed in the Naples setup due to the use of an old version of the mechanical frames. As a consequence, a bending correction, described in Appendix \ref{sec:CORRECTION}, needs to be applied.

Only the \ac{DF}-\ac{XA}-\ac{SS} has been measured in Naples; the \ac{PDE} results are shown in Table~\ref{tab:PDE}. These values are computed as the weighted mean from the results of the source heights and analysis positions. 

\begin{table}[ht!]
    \centering
    \begin{tabular}{lc}
        \multicolumn{2}{c}{\acl{PDE}}          \\\hline
        \ac{OV} (\unit{\volt})     & INFN Naples \\ \hline \hline
        \num{4.5} & \SI{3.1(5)}{\percent}      \\ 
        \num{6.0} & \SI{3.6(5)}{\percent}      \\ 
        \num{7.0} & \SI{4.0(6)}{\percent}      \\ \hline

    \end{tabular}
    \caption{\ac{PDE} as a function of \ac{OV} for the baseline \ac{DF}-\ac{XA}-\ac{SS} in the Naples setup.}
    \label{tab:PDE}
\end{table}

The large uncertainty attributed to the bending correction, \SI{\sim14}{\percent} (see Appendix~\ref{sec:CORRECTION}), has led to a larger uncertainty in the final \ac{PDE} value of \SI{\sim15}{\percent}.



\subsection{\acs{XA} \acs{PDE} Comparison\label{sec:RESULTS_COMPARISON}}
\begin{figure}[ht!]%
    \centering
    \includegraphics[width=1\linewidth]{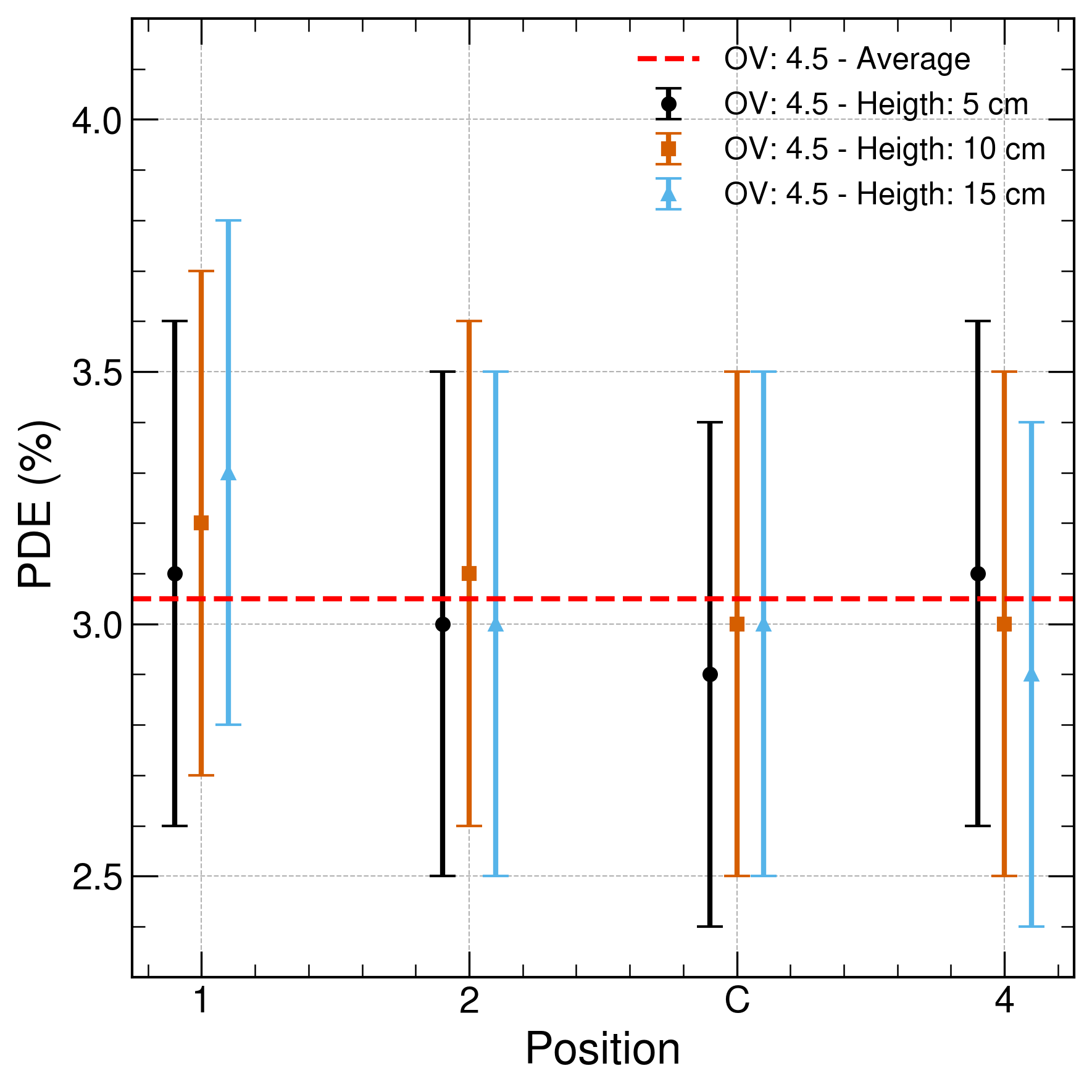} %
    \caption{Baseline \ac{XA} \ac{PDE} from the simulation method. Showing example source height scan for \SI{4.5}{\volt} overvoltage.}%
    \label{fig:PCE1}%
\end{figure}

The \ac{PDE} measurements of the baseline \ac{DF}-\ac{XA}-\ac{SS} model are shown in Figure~\ref{fig:box_pde} and Figure~\ref{fig:PCE1} for the reference and simulation methods, respectively, as a function of the alpha source position relative to the \ac{XA}. 

As can also be seen in Tables~\ref{tab:results} and~\ref{tab:PDE}, these results show a minor but consistent difference (\SI{\sim16}{\percent}) between the two methods, which can be attributed to the limitations of the bending correction for the simulation method presented in Appendix \ref{sec:CORRECTION}. This correction was not required for \ac{CIEMAT}'s measurements thanks to the improved \ac{XA} mechanical frame, which ensures the alignment between the \ac{WLS--LG} and the \acp{SiPM} in all positions.



Both the reference and simulation methods show the expected increase in \ac{PDE} for source positions that are closer to the device's edges, where the \acp{SiPM} are placed. 

\section{Conclusions\label{sec:CONCLUSION}} 
The work presented in this publication provides a comprehensive analysis of the performance of the \ac{XA} proposed for the \ac{DUNE} \ac{FD} \ac{VD} \ac{PDS}. It highlights the comparison of various \ac{XA} configurations, allowing the \ac{DUNE} collaboration to make an informed decision on the final choice of the \ac{XA} device and continue the design development for future detectors.

We have proven the compatibility, reproducibility and reliability of the results by using two different methods and setups for the \ac{PDE} measurement. The reference method has provided an absolute \ac{PDE} of \SI{3.7(3)}{\percent} and the simulation method a value of \SI{3.1(5)}{\percent} at \ac{OV} \SI{4.5}{\volt} for the \ac{SS} version of the baseline design. These results are compatible within 1$\sigma$ and confirm that the device meets the minimal requirement set by \ac{DUNE}'s physics goals.

Having validated the results of the reference method, we present additional measurements carried out at \ac{CIEMAT}.
We demonstrate the improvement in \ac{PDE} for \ac{XA} configurations without \ac{DF}, as expected from simulations. The measured increase is \SI{18}{\percent} and \SI{11}{\percent} for \ac{SS} and \ac{DS} configurations, respectively. The obtained \ac{PDE} value is \SI{4.5(4)}{\percent} at \SI{4.5}{\volt} \ac{OV}. Therefore, we recommend removing the \ac{DF} from the baseline \ac{XA} in \ac{DUNE} \ac{FD} \ac{VD} design.


We observe no remarkable difference for the \ac{WLS--LG} of \SI{24}{\milli\gram\per\kilo\gram} chromophore concentration and \SI{5.5}{\milli\meter} width with respect to the baseline \ac{XA}. Nevertheless, this configuration does have a flatter \ac{PDE} across the \ac{XA} surface. We conclude that the simultaneous modification of dye concentration and \ac{WLS--LG} thickness requires further optimisation. 

We also show that the \ac{XA}-\ac{SS} and \ac{DS} models have similar efficiencies. This suggests that most of the photons that escape the \ac{WLS--LG} are not collected.

The absolute \ac{PDE} of the \ac{XA} is a crucial parameter for the \ac{DUNE} \ac{PDS}, as it directly influences the detector's sensitivity to \ac{VUV} light and, consequently, its ability to accurately reconstruct neutrino interactions. The results presented in this publication have been instrumental in optimising the proposed design for the XA and provide valuable insights for future \ac{DUNE} \ac{PDS} developments.


\begin{appendices}
\section{Geant4 Simulations\label{sec:SIMULATION}}
\subsection{Reference Method\label{sec:CIEMAT_SIMULATION}}
The characterisation box at the reference setup has been designed using a Geant4 simulation with the goal of avoiding the shadowing of the inner surfaces, allowing light produced by the alpha particles to propagate directly to the different sensitive areas.
The parameters of the characterisation box are presented in Table \ref{tab:dimensions}. The distance to the source is computed from centre-to-centre in 3D; the effective area of each sensor is defined as the inner surface with geometrical acceptance for light from the source; the number of collected photons is defined as the number of photons simulated to arrive at the effective surfaces; and finally, the average incidence angle is the average angle between the incoming photon vector and the sensor's plane (in this definition, \ang{90} corresponds to a perpendicularly impinging photon).

\begin{table}[htp]
\centering
    \resizebox{\columnwidth}{!}{
    	\begin{tabular}{lccc}
            \multicolumn{4}{c}{Characterisation Box Simulation }  \\
    	    \hline
            Parameters                                   & Input/Output & \ac{XA}     & Ref. \ac{SiPM} \\ \hline \hline
            Distance to Source (\unit{\milli\meter})     & input        & \num{74}    & \num{29}       \\
            Effective area (\unit{\square\milli\meter})  & input        & 145.5 $\times$ 145.5 & 6 $\times$ 6       \\
            Collected Photons                            & output       & \num{25900} & \num{603}      \\ 
    	    Average Incidence Angle (\unit{\degree})     & output       & \ang{90}    & \ang{75}       \\ \hline
        \end{tabular}
    }
	\caption{Simulation parameters from the characterisation box simulation. See text for details.} 
	\label{tab:dimensions}
\end{table}

\begin{figure}[ht!]
    \centerfloat
    \includegraphics[width=1.3\linewidth]{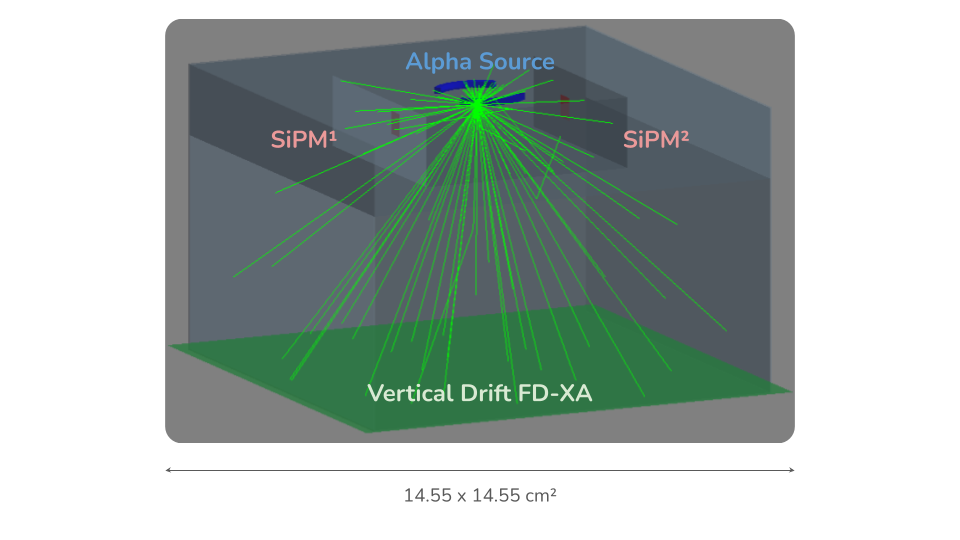}
	\caption{\ac{MC} simulation of the characterisation box tracking 100 photon trajectories.}
	\label{fig:MC}
\end{figure}

Figure~\ref{fig:MC} shows the simulated geometry. The walls of the characterisation box are 3D-printed from a black plastic material that fully absorbs the photons (shown here semi-transparently for illustration purposes). The source holder is displayed in blue. The two reference VUV4 \acp{SiPM} (red) and the \ac{XA} (green) are also highlighted in colours and configured as sensitive areas to retrieve the number of arriving photons. 

Instead of simulating alpha particles directly, the software generates exactly \num{195636}~$\gamma$ (corresponding to a light yield of \num{51000}~$\gamma$\unit{\per\mega\eV} and a quenching factor of \num{0.7}) and distributes them homogeneously over the source's radiative surface. 
These photons are propagated uniformly and isotropically across the LAr volume. The simulation provides the number of photons reaching the \ac{XA} acceptance window as well as each of the reference SiPM sensors, which resulted in \numlist{25900(240);603(7)}~$\gamma$, respectively. The uncertainties are derived from a box tolerance study in which we iterated over different geometrical sizes and sensor positions within the estimated manufacturing errors. Additionally, the resulting data provides the average incidence angle of all collected photons. This is used to correct the reference \ac{SiPM} \ac{PDE} measured in~\cite{Alvarez-Garrote:2024byb} by the angle dependence according to~\cite{nEXO:2019jhg}.
The geometrical correction factor of Equation~\ref{eq:ciemat_pde} is also evaluated from this simulation:

\begin{ceqn}
	\begin{equation}
		f^\text{geo} = \dfrac{\text{MC}^\text{photons}_\text{SiPM}}{\text{MC}^\text{photons}_\text{XA}} = 0.0465 \pm 0.0007 \,.
		\label{eq:geometrical}
	\end{equation}
\end{ceqn}

\subsection{Simulation Method\label{sec:NAPLES_SIMULATION}}
To estimate the \ac{PDE} of the \ac{XA}, it is necessary to compare the number of detected \acp{PE} with the number of photons produced by scintillation reaching the sensor's optical windows. The key difference between this method and the previously presented reference method is that this simulation is required to provide the absolute number of true photons rather than a relative value. This may be less robust due to the uncertainties in the optical properties of the components at \ac{CT} and their implementation in a \ac{MC} simulation. The photon count was estimated using a Geant4 simulation. 
Photons were generated isotropically from the source and propagated through the \ac{LAr} medium until they reached the \ac{XA}. 
The different positions of the alpha source used during the measurement (four positions in the plane and three different heights) were simulated. To estimate the error, the source position was varied by \SI{0.5}{\centi\meter} in x, y and z directions. From the simulation, one can extract the expected response of the \ac{XA} in terms of photon collection. 

\begin{figure}[ht!]%
    \centering
    \includegraphics[width=\linewidth]{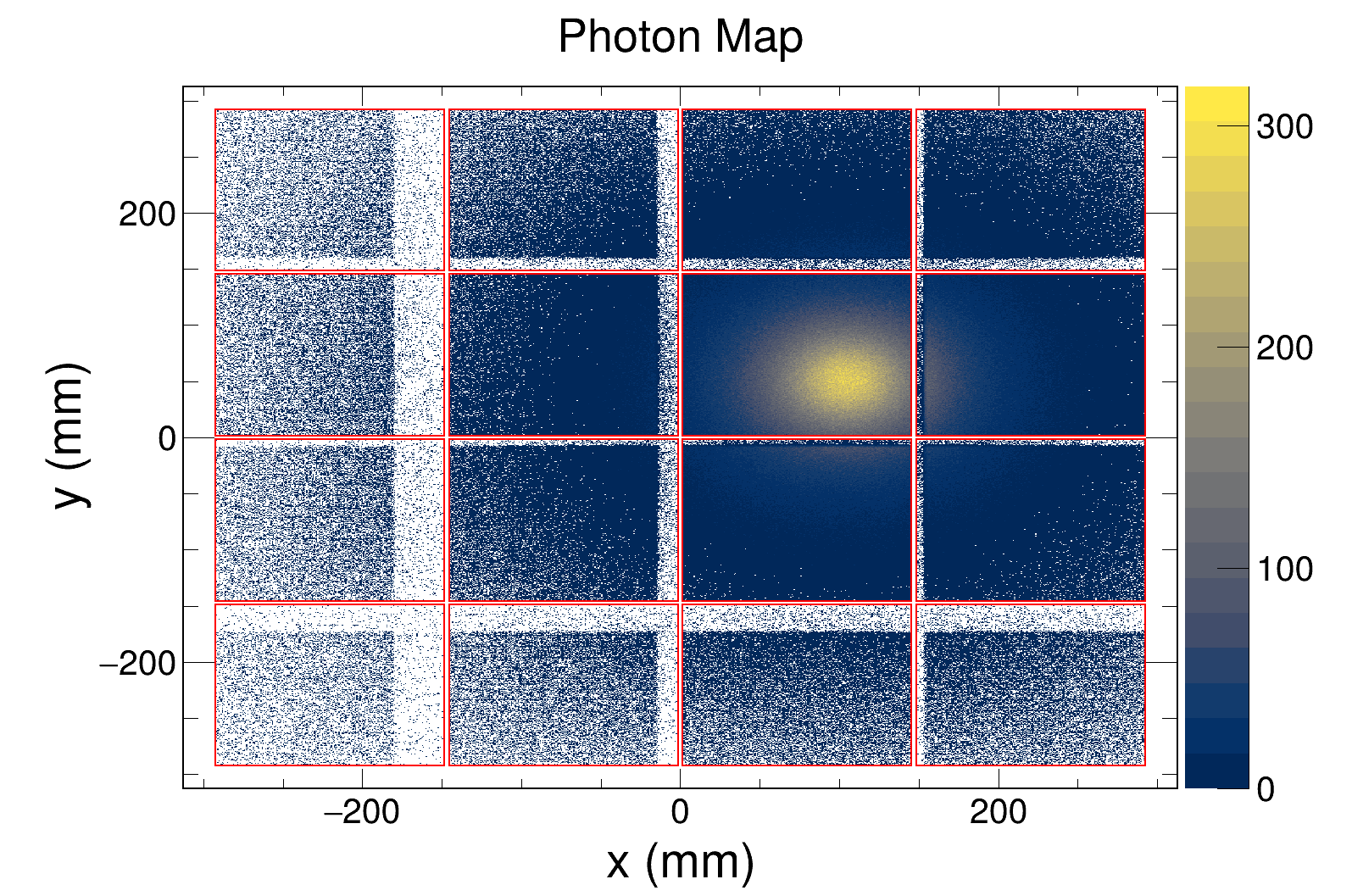} %
    \caption{Mapping of the photons arriving at the \ac{XA}, where the red squares are the optical windows.}
    \label{fig:geant4_simulation}%
\end{figure}

Figure~\ref{fig:geant4_simulation} shows the mapping of the photons arriving at the \ac{XA} optical windows, indicated by red squares. The source is located at position 2 (see Appendix~\ref{sec:NAPLES_SETUP}) at a height of 5 cm. The presence of the shadow region is due to the photons coming from the source hitting the grids of the device. For the source placed 5 cm away, the ratio between photons hitting the grid and photons reaching the optical window is approximately \SI{15}{\percent}.

\begin{figure}[ht]%
    \centering
    \includegraphics[width=\linewidth]{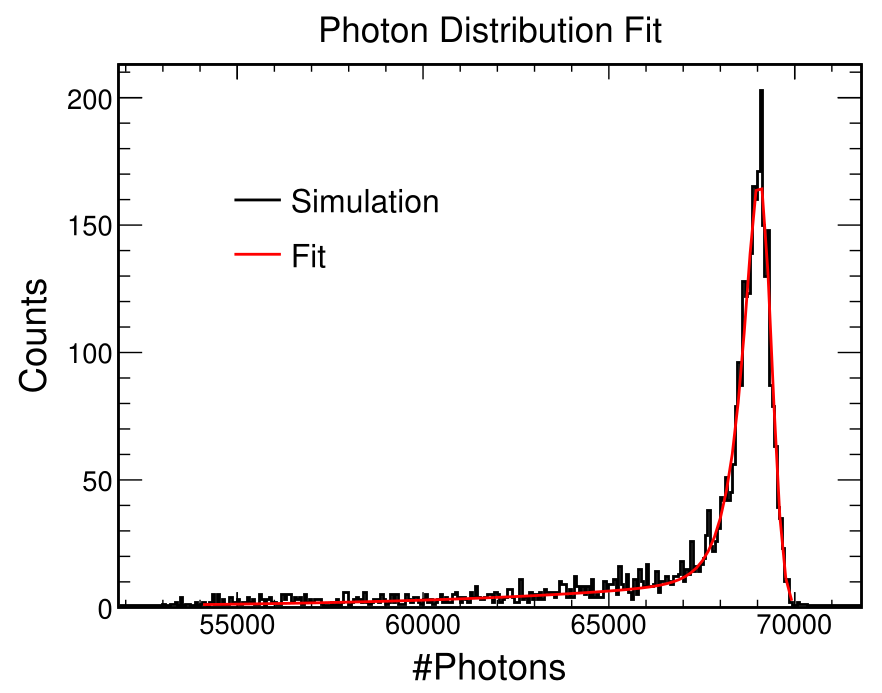} %
    \caption{Example of the number of photons reaching the optical windows.}
    \label{fig:geant4_charge}%
\end{figure}

Equation~\ref{eq:alpha_fit} is used to fit the charge spectrum of the photons reaching the XA optical windows~\cite{ref:alpha_peak}, as illustrated in Figure~\ref{fig:geant4_charge}.



\section{Bending correction\label{sec:CORRECTION}}

\begin{figure}[ht!]
    \centering
    \includegraphics[trim={5.5cm 0cm 5.5cm 0cm},clip, width=\linewidth]{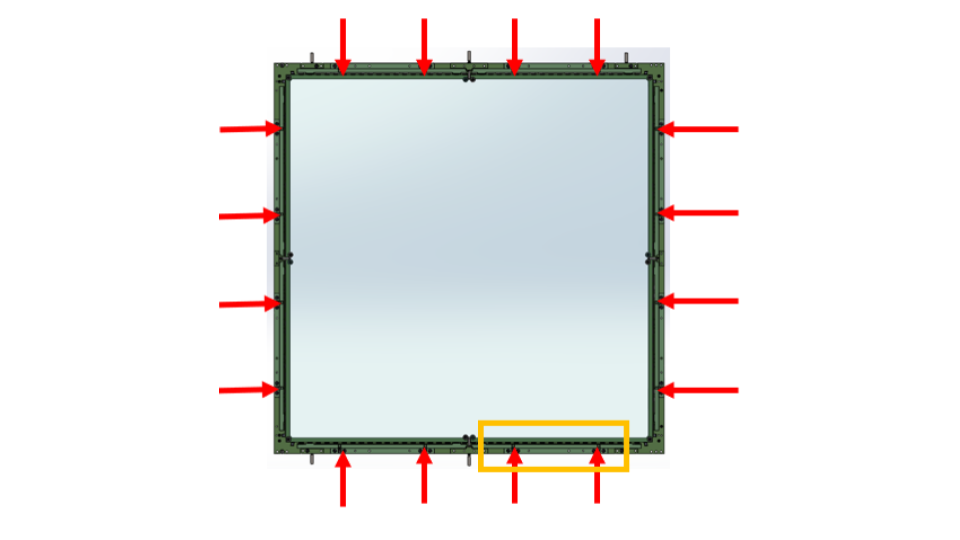} \\
    \vspace{-1cm}
    \includegraphics[width=\linewidth]{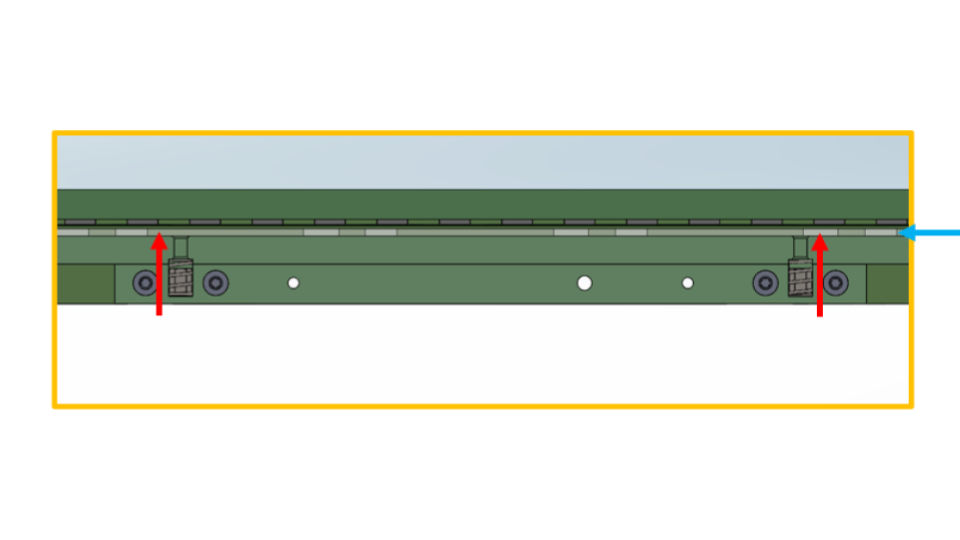}
    \vspace{-1cm}
    \caption{Schematic representation of the \ac{VD} \ac{XA} spring-loaded mechanism. \acp{SiPM} are mounted between the \ac{WLS--LG} and an acrylic interface strip (blue arrow). Springs (red arrow) apply constant pressure orthogonal to this interface. Thermal shrinkage at cryogenic temperatures of the \ac{WLS--LG} is \SI{6}{\milli\metre}, which causes the springs to partially decompress to \SI{11.5}{\milli\metre}. These springs are mechanically constrained to exert force and translate solely orthogonal to the \ac{WLS--LG} face. At each side of the frame, the centre position of the \ac{XA} is reinforced by a clamping bracket.}
    \label{fig:xa_spring_system}
\end{figure}

A spring-loaded mechanism ensures optimal contact between the \ac{SiPM} active surfaces and the \ac{WLS--LG} edges, counteracting the cryogenic contraction of the system. With respect to the design described in \ac{DUNE}'s \ac{TDR}~\cite{DUNE:2023nqi}, the mechanical frame of the \ac{XA}'s baseline design underwent some modifications. Most noticeably, four clamping brackets were added to the frame to mitigate the \ac{WLS--LG} misalignment under bending conditions, which can occur during horizontal assembly, storage, or transportation (see Figure~\ref{fig:xa_spring_system}). This change was not implemented in the \ac{XA} prototype measured by the \ac{INFN} Naples team, requiring a bending correction in the calculation of the PDE (see Section~\ref{sec:NAPLES_PDE}). The resulting misalignment between the \ac{WLS--LG} and the \acp{SiPM} was not present in \ac{CIEMAT}'s measurements, thanks to the aforementioned mechanical improvements and the vertical orientation during data taking. 

During the inspection phase of the Naples setup, it was discovered that the light guide had bent and there was a significant misalignment between the light guide plane and the \ac{SiPM} flex boards. This issue can be attributed to the fact that during the measurement phase, the \ac{XA} was positioned horizontally, with the light guide supported on only two sides rather than all four. As this was one of the first \ac{XA} prototypes, the problem observed in this test led to the inclusion of additional support points for the light guide in the design, which were improved in subsequent devices. The emptying of the vessel was performed in a slow and controlled manner, minimizing the likelihood of inducing any additional mechanical stress or deformation during this phase.

\begin{figure}[ht!]
    \centering
    \includegraphics[width=\linewidth]{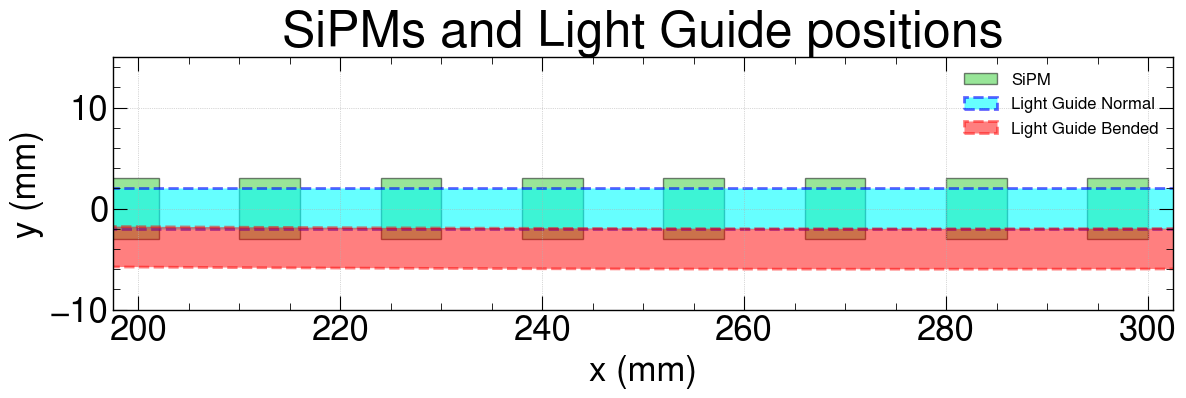}
    \includegraphics[width=\linewidth]{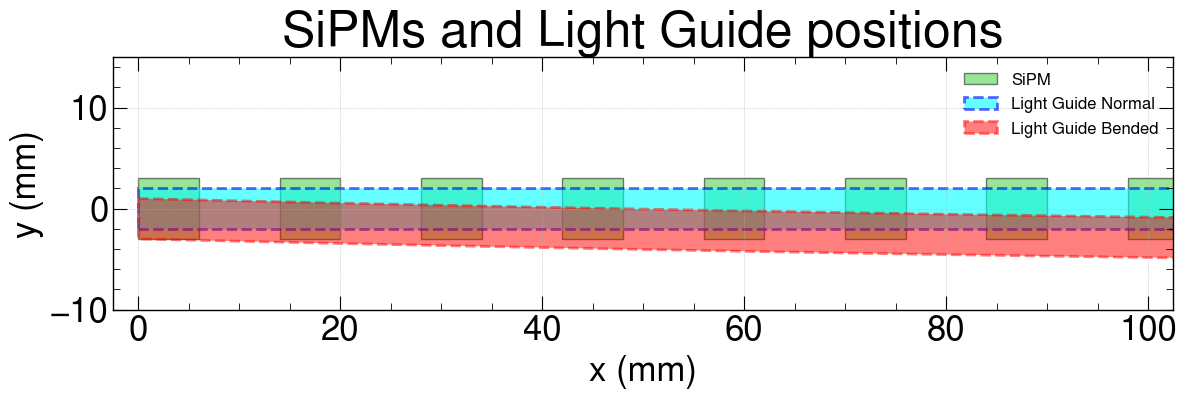}
    \caption{Schematic of the bent \ac{XA} light guide. The upper panel shows the central region of the light guide, while the lower panel shows the edge region. The green squares represent the internal \acp{SiPM}, each with a side length of \SI{6}{\milli\metre} and a centre-to-centre spacing of \SI{14}{\milli\metre}. The blue region represents the unbent light-guide configuration, while the red region represents the bent light-guide configuration.}
    \label{fig:bending1}
\end{figure}

To account for this issue, the misalignment shifts of the \acp{SiPM} relative to the centre of the light guide along each side were carefully measured. A representation of the misalignment is shown in Figure~\ref{fig:bending1}, where the central and the edge part (top and bottom) of the \ac{XA} components are schematised. A vertical shift of approximately \SI{5}{\milli\metre} was observed at the midpoint, and a vertical offset of about \SI{2}{\milli\metre} was noted at the edges.

Due to the close placement of the \ac{SiPM} surfaces relative to the edges of the light guide, a correction factor was implemented to compensate for bending effects and adjust the photon collection efficiency. The active area of the SiPM and the WLS surface are comparable, so any mismatch between them can lead to significant inefficiencies. 

The bending was modeled using the catenary equation, which is a hyperbolic cosine function:

\begin{ceqn}
    \begin{equation}
        y=a\cdot\cosh\left(x/b\right)+c\,.
          \label{eq:hyperbolic_cosine}
    \end{equation}
\end{ceqn}

The ratio of the areas aligned with the \acp{SiPM} in the bent versus unbent light guide configurations was calculated to be $f^\text{bending}=\num{0.58(8)}$. The spread of the value comes from the uncertainty of the measurement of the displacement using a calibre (order of \SI{0.5}{mm}). It is important to note that this bending correction was characterised after the measurements, and might have been more dramatic during the data taking. Nevertheless, we provide an absolute \ac{PDE} value for the simulation method on the assumption that this correction holds and as a cross-check against the \ac{PDE} computed using the reference method.

\section{INFN Naples Setup}  \label{sec:NAPLES_SETUP}
The \ac{XA} \ac{PDE} was also measured at the Physics Department of the University of Naples Federico II and \ac{INFN}. The photodetector module was placed horizontally inside a cryostat capable of holding up to \SI{1}{\tonne} of LAr. 


The inner walls of the cryostat were lined with black Delrin material and hung on a stainless steel thermal shield, improving optical insulation and preventing the reflection of photons from the stainless steel wall of the cryostat. With the same purpose, a Delrin disk was placed at the bottom side of the shield. The cryostat was filled with \ac{LAr} with a level of contamination of the order of $\sim$5~ppm ($<$ 1.5 ppm N$_{2}$, $<$ 1 ppm H$_{2}$O, $<$ 1 ppm O$_{2}$ and others). The \ac{LAr} was filtered using an in-line Trigon (Engelhard Q5--Cu0226) purifier, reducing the amount of oxygen diluted in the \ac{LAr} to the level of tenths of ppb or less. Before filling, the cryostat was pumped to achieve a vacuum of \SI{\sim 1e-3}{\milli\bar} after some cycles of pumping and purging with gaseous argon to reduce the contribution of outgassing. The cryostat was filled up to a height of \SI{20}{\centi\metre} above the \ac{XA}.

\begin{figure}[ht!]
    \centering
    \includegraphics[width=\linewidth]{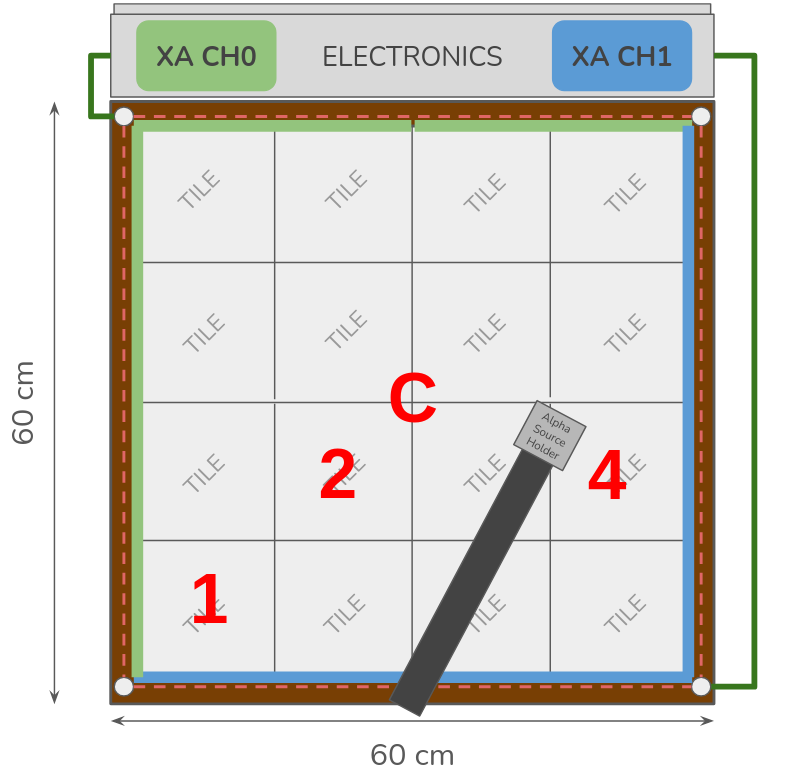}
    \caption{\ac{XA} channel arrangement in relation to the four reference positions of the alpha source.}
    \label{fig:channels}
\end{figure}

To monitor the liquid argon purity, a TPB-coated photomultiplier tube from Hamamatsu R11065 (\href{https://www.hamamatsu.com}{HPK}) was used to measure the slow component of the liquid argon scintillation time profile. An optical fibre was used to drive pulsed laser flashes inside the cryostat to calibrate the \ac{XA} module. A \SI{250}{\becquerel} activity $^{241}$Am source was placed above the \ac{XA} in an arm connected to a motion feedthrough, allowing the source to be changed in position with respect to the \ac{XA}. The motion system was able to vary linearly the distance from the \ac{XA} and rotate the arm where the source was placed, providing to radiate a large portion of the photosensor (see Figure~\ref{fig:channels} and Table~\ref{tab:alpha_position}). The test is done in different positions and heights to be sure that the efficiency is almost independent of the position. Alpha particles decaying from the $^{241}$Am produced enough scintillation light for the final evaluation of the \ac{PDE}. 
The data were digitised using a \href{https://www.caen.it/products/dt5725/}{CAEN} V1725, with a 14-bit ADC and a sampling frequency of \SI{250}{\mega\hertz}. Each channel reads two sides of the \ac{XA} and a schematic can be seen in Figure~\ref{fig:channels}.

 \begin{table}[ht!]
    \centering
   
    \small 
    \begin{tabular}{lcc}
    \multicolumn{3}{c}{Measurement Source Positions} \\
    \hline
    Position & x (\unit{\centi\meter}) & y (\unit{\centi\meter}) \\ \hline \hline
    1        & 22     & 21     \\
    2        & 10.5   & 5      \\
    C        & 0      & 0      \\
    4        & -22.5  & 3      \\
    \hline
    \end{tabular}
    \caption{Spatial coordinates of each position of the alpha source in relation to the XA centre as implemented for the simulation method in Naples' setup.}
     \label{tab:alpha_position}
\end{table}

\end{appendices}

\begin{acknowledgements}
The present research has been supported and partially funded by MCIN/AEI/10.13039/ 501100011033 under Grants no. PID2023-147949NB-C51, No. PID2023-147949NB-C52, CNS2023--144183 and PRE2020–094863 of Spain, by the Italian Ministero dell’Università e della Ricerca (PRIN 2017KC8WMB and PRIN 20208XN9TZ), by the European Union’s Horizon 2020 Research and Innovation programme under Grant Agreement No 101004761 (AIDAinnova), by the European Union-Next Generation EU, and finally by the University of Ferrara (FIR2023), the BiCoQ Center of the University of Milano Bicocca and 
Leonardo Grant for Researchers in Physics 2023 BBVA Foundation, LEO23--1--9021.

\end{acknowledgements}

\bibliography{references}


\begin{acronym}
  \acro{APA}{Anode Plane Assembly}
  \acro{BSM}{Beyond the Standard Model}
  \acro{CPA}{Cathode Plane Assembly}
  \acro{CC}{Charged Current}
  \acro{CIEMAT}{Centro de Investigaciones Energéticas Medioambientales y Tecnológicas}
  \acro{CERN}{European Organization for Nuclear Research}
  \acro{CT}{Cryogenic Temperature}
  \acro{DAQ}{Data Acquisition}
  \acro{DF}{Dichroic Filter}
  \acro{DS}{Double--Sided}
  \acro{DUNE}{Deep Underground Neutrino Experiment}
  \acro{ES}{Elastic Scattering}
  \acro{FD}{Far Detector}
  \acro{Fermilab}{Fermi National Accelerator Laboratory}
  \acro{HD}{Horizontal Drift}
  \acro{INFN}{Istituto Nazionale di Fisica Nucleare}
  \acro{LAr}{Liquid Argon}
  \acro{LN$_2$}{Liquid Nitrogen}
  \acro{GAr}{Gaseous Argon}
  \acro{LArSoft}{Liquid Argon Software Framework}
  \acro{LArTPC}{Liquid Argon Time Projection Chamber}
  \acro{LY}{Light Yield}
  \acro{MARLEY}{Model of Argon Reaction Low Energy Yields}
  \acro{MC}{Monte Carlo}
  \acro{ML}{Machine Learning}
  \acro{NC}{Neutral Current}
  \acro{PCE}{Photon Collection Efficiency}
  \acro{PDE}{Photon Detection Efficiency}
  \acro{PDS}{Photon Detection System}
  \acro{PE}{Photoelectron}
  \acro{PMMA}{Polymethyl Methacrylate}
  \acro{PMT}{Photomultiplier Tube}
  \acro{pTP}{p--Terphenyl}
  \acro{OV}{Overvoltage}
  \acro{RT}{Room Temperature}
  \acro{RMS}{Root Mean Square}
  \acro{SiPM}{Silicon Photomultiplier}
  \acro{SPE}{Single Photoelectron}
  \acro{SS}{Single--Sided}
  \acro{STD}{Standard Deviation}
  \acro{SURF}{Sanford Underground Research Facility}
  \acro{TDR}{Technical Design Report}
  \acro{TPB}{Tetraphenyl Butadiene}
  \acro{TPC}{Time Projection Chamber}
  \acro{VD}{Vertical Drift}
  \acro{VUV}{Vacuum Ultraviolet}
  \acro{WLS--LG}{Wavelength Shifter Light Guide}
  \acro{XA}{X--ARAPUCA}
  \acro{XT}{cross-talk}
\end{acronym}


\end{document}